\documentclass[aapm,prb,preprint,graphicx,unsortedaddress,nofootinbib]{revtex4-1} %prb gives superscript references
\draft % marks overfull lines with a black rule on the right

\usepackage[mathlines]{lineno}% Enable numbering of text and display math
\usepackage{graphicx}
\usepackage{color}

\usepackage{amssymb} %For \mathbb
\usepackage{amsmath} %For \align
\usepackage{bm} %For \bm
\usepackage{upgreek}
\usepackage{longtable}

\newcommand{\ve}[1]{\mathbf{#1}} %Vector in math mode. note that I had to use bm instead for some greek letters!
\newcommand{\ma}[1]{\mathsf{#1}} %Matrix in math mode. Try this and see if it improves clarity!
\newcommand{\imu}{i} %Imaginary unit %If we want to use another style for the imaginary unit
\newcommand{\trsp}{^{\mathsf{T}}} %transpose
\newcommand{\Rtwo}{{\mathbb{R}^2}}
\newcommand{\fourop}{\mathcal{F}} %May want to change to mathsf for consistency with matrices?s

\newcommand{\ch}[1]{\textcolor{black}{#1}}
\newcommand{\chtwo}[1]{\textcolor{black}{#1}}

\let\oldequation\equation
\let\oldendequation\endequation

\renewenvironment{equation}
  {\linenomathNonumbers\oldequation}
  {\oldendequation\endlinenomath}

\let\oldalign\align
\let\oldendalign\endalign

\renewenvironment{align}
  {\linenomathNonumbers\oldalign}
  {\oldendalign\endlinenomath}

\begin{document}
\title{A framework for performance characterization of energy-resolving photon-counting detectors}

\author{Mats Persson}
\email[Author to whom correspondence should be addressed. Electronic mail: ]{matspers@stanford.edu}
\affiliation{Departments of Bioengineering and Radiology, Stanford University, Stanford 94305, California, USA\\ (Corresponding author. E-mail: matspers@stanford.edu)}
\author{Paurakh L. Rajbhandary}
\affiliation{Departments of Electrical Engineering and Radiology, Stanford University, Stanford 94305, California, USA}
\author{Norbert J. Pelc}
\affiliation{\ch{Departments of Bioengineering, Radiology and Electrical Engineering}, Stanford University, Stanford 94305, California, USA}

\date{\today}

\begin{abstract}
%\boldmath
\textbf{Purpose}: Photon-counting energy resolving detectors are subject to intense research interest, and there is a need for a general framework for performance assessment of these detectors. The commonly used linear-systems theory framework, which measures detector performance in terms of noise-equivalent quanta (NEQ) and detective quantum efficiency (DQE) is widely used for characterizing conventional X-ray detectors but does not take energy-resolving capabilities into account. The purpose of this work is to extend this framework to encompass energy-resolving photon-counting detectors and elucidate how the imperfect energy response \ch{and other imperfections in} real-world detectors \ch{affect} imaging performance, \ch{both for feature detection and for material quantification tasks.} 
%Discarded: \ch{In addition, we demonstrate that the new framework generalizes the Cram\'{e}r-Rao lower bound and thereby provides a method of assessing detector performance in material quantification tasks}.\\

\textbf{Method}: We generalize NEQ and DQE to matrix-valued quantities as functions of spatial frequency, and show how these matrices can be calculated from simple Monte Carlo simulations. To demonstrate how the new metrics can be interpreted, we compute them for simplified models of fluorescence and Compton scatter in a photon-counting detector and for a Monte Carlo model of a CdTe detector with $0.5\times 0.5 \; \mathrm{mm}^2$ pixels.\\
\textbf{Results}: Our results show that the ideal-linear-observer performance for any detection or material quantification task can be calculated from the proposed generalized NEQ and DQE metrics. \ch{We also demonstrate that the proposed NEQ metric is closely related to a generalized version of the Cram\'{e}r-Rao lower bound commonly used for assessing material quantification performance.} Off-diagonal elements in the NEQ and DQE matrices are shown to be related to loss of energy information due to imperfect energy resolution. The Monte Carlo model of the CdTe detector predicts a zero-frequency dose efficiency relative to an ideal detector of 0.86 and 0.65 for detecting water and bone, respectively. When the task instead is to quantify these materials, the corresponding values are 0.34 for water and 0.26 for bone.\\
\textbf{Conclusions}: We have developed a framework for assessing the performance of photon-counting energy-resolving detectors and shown that the matrix-valued NEQ and DQE metrics contain sufficient information for calculating the dose efficiency for both detection \ch{and} quantification tasks, the task having any spatial and energy dependence. This framework will be beneficial for the development and optimization of photon-counting X-ray detectors.
\end{abstract}

% insert suggested PACS numbers in braces on next line

\maketitle %\maketitle must follow title, authors, abstract and \pacs

% Body of paper goes here. Use proper sectioning commands. 
% References should be done using the \cite, \ref, and \label commands

%Papers I thought of citing but in the end decided not to: cite barrett Objective assessment of image quality. II.    Fisher information, Fourier crosstalk, and figures of merit for task performance, Rigle and Lariviere 10.1109/NSSMIC.2012.6551856
\section{Introduction}
\label{sec:introduction}
The development of energy-resolving photon-counting detectors for medical X-ray imaging is an active research field.\cite{taguchi_vision_2020,shikhaliev_pcct_concept_initial_results, bornefalk_silicon_strip_pmb, shikhaliev_first_results,giersch_energy_weighting} Systems with such detectors are currently commercially available for mammography\cite{berglund_energy_weighting}, and prototypes have been demonstrated for computed tomography (CT)\cite{schlomka_experimental_feasibility, persson_energy_resolved_ct_imaging, yu_mayo_research_prototype_evaluation, ronaldson_toward_quantifying}. An energy-resolving photon-counting detector uses several electronic threshold levels to separate the registered counts into energy bins depending on the deposited energy \ch{from} each event. This enables spectroscopic imaging, i.e. using the information contained in the energy distribution of the incoming spectrum to obtain more information about the imaged object. The energy information can be used for improving the the signal-difference-to-noise ratio (SDNR) by optimal weighting\cite{cahn_dqe, TGS_image_based}, for removing beam-hardening artifacts and for generating material-selective images.\cite{alvarez_macovski,roessl_proksa_k_edge,schlomka_experimental_feasibility}

The ongoing development of improved energy-resolving photon-counting X-ray detectors raises the question of how detector performance should be measured. Using a relevant metric is important in order to optimize and compare detector designs \ch{comprising} different detector materials and pixel sizes or with different properties of the readout electronics such as the energy resolution and anti-coincidence logic.

The performance of conventional, non-energy-resolving X-ray detectors, such as energy-integrating detectors and photon-counting detectors without energy-resolving capabilities, is commonly measured in terms of the detective quantum efficiency (DQE).\cite{icru_54} This metric, which is \ch{ideally} expressed as a function of spatial frequency, measures how well the detector is able to use the incoming photon statistics for each spatial frequency compared to an ideal detector. A closely related metric is the noise-equivalent quanta (NEQ), which measures the number of photons that an ideal detector would require in order to obtain the same signal-difference-to-noise ratio as the studied detector. An overview of the theory of DQE and NEQ for ordinary (non-spectral) detectors can be found in Ref. \citenum{cunningham_linear_systems_chapter}.

Several studies have been \ch{published on} how to model the frequency-dependent detective quantum efficiency of photon-counting detectors,\cite{acciavatti_maidment_phc_ei,tanguay_dqe_2013,xu_cascaded_sys_phc, stierstorfer_dqe} but these studies do not take energy information into account. \ch{At the same time}, the impact of detector nonidealities on spectral imaging tasks has been studied by several authors\cite{chen_optimization_of_beam_q,cammin_evaluation_spectral_dist,taguchi_energy_response_pileup,tanguay_dqe_2015,faby_statistical_correlations,taguchi_spatioenergetic,rajbhandary_spatioenergy_spie}. However, these publications only study the zero-spatial-frequency performance and do not investigate the frequency-dependent spectral performance. Shikhaliev et al. \cite{shikhaliev_characteristic_x_ray} studied the effect of characteristic X-rays on the spectrum shape and on the spatial resolution. Richard and Siewerdsen\cite{richard_noise_reduction_dual_energy} used linear-systems theory to study the performance of a dual-energy system. Also, the detectability for specific tasks with both spectral and spatial dependence has been studied by Fredenberg et al.\cite{fredenberg_contrast_enhanced_spectral}, Yveborg et al. \cite{yveborg_theoretical_comp} and Chen et al.\cite{chen_size_dependent_parameters}
%Removed yun_csa_singlez since it seems to model energy integrating detectors

A common way to combine information from several spectral channels is to make a weighted sum of the energy-selective images, with weights chosen to be optimal for the considered imaging task.\cite{tapiovaara_wagner,cahn_dqe,TGS_image_based} A good measure of the performance of an energy-resolving detector should therefore take optimal weighting into account. Taking the spatial-frequency dependence of the signal and noise into account when calculating optimal weights can improve detectability in an energy-weighted image, either by calculating an optimized weight for each image\cite{bornefalk_task_based} or by letting the weights themselves be functions of spatial frequency\cite{yveborg_frequency_based_weighting}. This is equivalent to applying different spatial filters to the different energy bin images before forming the weighted sum, an approach studied in the dual-energy case by Richard and Siewerdsen.\cite{richard_noise_reduction_dual_energy}

Another way to utilize spectral X-ray \ch{measurements} is to perform basis material decomposition, which builds on the observation that the linear attenuation coefficient of any substance in the human body can be approximated well as a linear combination of a small number of basis functions, typically two \ch{in the absence and} three in the presence of a high-atomic-number contrast agent with a K-edge in the diagnostic energy range.\cite{alvarez_macovski} Basis material decomposition then amounts to using the energy-resolved X-ray data to estimate the amount of each of these basis materials along each projection line, thereby completely characterizing the energy-dependent attenuation of the object.\cite{alvarez_macovski,roessl_proksa_k_edge} This is useful in particular for CT, to remove beam-hardening artifacts and characterize object composition. \ch{Assessment of the potential performance of X-ray detectors for basis material decomposition is typically made using the Cram\'er-Rao lower bound (CRLB), which gives a lower bound for the variance in a basis material image estimated from the measured data. Most previous studies of detector performance utilizing the CRLB have studied only the zero-frequency (large-area) performance, although an extension to higher spatial frequencies has been published recently.\cite{rajbhandary_frequency_dependent_dqe}}

The purpose of this work is to \ch{extend and unify the above-mentioned methods of detector characterization} into a general framework for comparing the performance of energy-resolving imaging detectors, encompassing both energy-weighting and material-decomposition performance. To this end, we derive an expression for the detectability achieved by an ideal linear observer, taking into account the full spatial-frequency dependence of the available energy information. We also propose generalized NEQ and DQE metrics for measuring energy-resolving detector performance. \ch{Furthermore, we study the performance attainable in material quantification tasks by deriving an expression for the frequency-dependent CRLB, and show that the CRLB is closely related to the proposed generalized NEQ metric. As with any NEQ and DQE formalism, we assume a linear system, although the relevance to situations where nonlinear imaging is used is discussed.} \ch{The outline of the article is as follows: In Sec. \ref{sec:theory}, we  derive the expressions for detectability, NEQ, DQE and CRLB and show how they are related to each other.} To show how the proposed metrics can be used in practice, \ch{we show in Sec. \ref{sec:materials_and_methods} how this framework can be applied to two} simple models of detectors with nonideal energy response, and to a more realistic simulation model of a photon-counting detector. \ch{In Sec. \ref{sec:results} we present the resulting NEQ and DQE metrics for each of these three models, and we discuss the implications of these results in Sec. \ref{sec:discussion} and \ref{sec:conclusion}.} 

%Discarded text
%As it turns out, the resulting metrics are not scalar quantities but matrix-valued, since different energies can affect each other in the result. 
\section{Theory}
\label{sec:theory}
\subsection{Matrix-valued NEQ and DQE}

In the following, \ch{we use} parentheses to denote that a quantity is a function of a continuous variable \ch{and subscripts} when the variable is discrete. Most of our notation is similar to Ref. \citenum{cunningham_linear_systems_chapter} and \citenum{tanguay_dqe_2013}. Consider a situation where a distribution of X-ray photons is incident on a detector. Let $q(\ve{r},E)$ be a random process describing the number of incident photons per area and energy as a function of position $\ve{r}=(x,y)\trsp$ and energy $E$. This photon distribution is measured by an energy-resolving detector and registered as counts in energy bins $d^s_{\ve{n},k}$ where $\ve{n}=(n_x,n_y)\trsp$ is the discrete coordinate of a detector pixel in the $(x,y)$ plane and $k=1,\ldots,N_b$ is the index of the energy bin. To avoid boundary effects, we assume that the detector and the photon distribution are of infinite extent. Both $q$ and $d^s$ are random processes, \ch{with} expected \ch{values $\overline{q}$} and $\overline{d}^s$, respectively. It is also useful to define a presampling detected signal $d_k(\ve{r})$, whose value at every point $\ve{r}$ is the number of counts that would be registered in each energy bin $k$ by a fictitious pixel centered at $\ve{r}$. Therefore, $d^s_{\ve{n},k}=d_k(\ve{r_n})$ where $\ve{r_n}=(n_x \Delta_x, n_y \Delta_y)\trsp$ is the center of pixel $\ve{n}$ and $\Delta_x$ and $\Delta_y$ are the pixel center-to-center distances in the $x$ and $y$ directions, respectively.

The continuous Fourier transform of $q(\ve{r},E)$ with respect to $\ve{r}$ is a random process given by $Q(\ve{u},E)=\mathcal{F}\left[q(\ve{r},E)\right](\ve{u},E)=\int_{\Rtwo} q(\ve{r},E)\mathrm{e}^{-2\pi \imu\ve{u}\cdot\ve{r}} \mathrm{d}\ve{r}$, and $D_k(\ve{u})$ is defined analogously as the transform of $d_k(\ve{r})$. The 2D discrete-space Fourier transform of $d^s_{\ve{n},k}$ with respect to $\ve{n}$ is given by $D^s_k(\ve{u})=\fourop_{\mathrm{DS}}\left[d^s_{\ve{n},k} \right](\ve{u})=\sum_{\ve{n}=-\infty}^{\infty} d^s_{\ve{n},k}\mathrm{e}^{-2\pi \imu\ve{u}\cdot\ve{r_n}}$. We now assume that the system is linear and shift-invariant, which allows us to introduce the system point-spread function for the $k^{\mathrm{th}}$ energy bin $h_k(\ve{\Delta r},E)$ that relates the expected values of the the incoming photon distribution and the registered count distribution:
\begin{equation}
  \overline{d}_k(\ve{r})=\Delta_x\Delta_y\int_\Rtwo\int_{0}^{E_{\mathrm{max}}}h_k(\ve{r}-\ve{r}',E) \overline{q}(\ve{r}',E) \mathrm{d}E\mathrm{d}\ve{r}',
\end{equation}
%The notation is admittedly a bit fuzzy here. qbar here means expected incodent quanta in general, but later on it means expected incident quanta in the background case.

where $E_{\mathrm{max}}$ is the maximum incident energy. Here, the pixel cell area $\Delta_x\Delta_y$ has been broken out of the definition of $h_k(\ve{\Delta r},E)$, meaning that $h_k$ can be interpreted as a presampling distribution of registered quanta per unit pixel area in the $k^{\mathrm{th}}$ energy bin when the input signal $\overline{q}(\ve{r},E)$ is a beam incident on a single point of the detector. $h_k(\ve{\Delta r},E)$ thus contains information about both the quantum detection efficiency and the \ch{spatial and energy} resolution of the detector. Denoting the Fourier transform of $h_k$ with respect to $\Delta_x$ and $\Delta_y$ by $H_k$,

\begin{equation}
  \overline{D}_k(\ve{u})=\Delta_x\Delta_y\int_{0}^{E_\mathrm{max}}H_k(\ve{u},E) \overline{Q}(\ve{u},E) \mathrm{d}E.
\end{equation}

In analogy with conventional X-ray detectors, it is possible to define a presampling modulation transfer function $\left[\mathrm{MTF_{pre}}\right]_k(\ve{u},E)=\left|H_k(\ve{u},E)/H_k(\ve{0},E)\right|$ for each incident energy and for each energy bin.

%Discarded, reinsert if needed: the gain $\overline{G}_k(E)=H_k(\ve{0},E)$
To calculate the performance of the detector for a given imaging task, we also need to know the covariance between different measurements $d^s_{\ve{n},k}$. Assuming that the noise is wide-sense stationary, this is given by the cross-covariance matrix $K^s$ with elements
\begin{equation}
  K^s_{\ve{\Delta n},k,k'}=\mathrm{Cov}(d^s_{\ve{n},k},d^s_{\ve{n}+\ve{\Delta n},k'}).
\end{equation}

The cross-spectral density of $d^s_{\ve{n},k}$ can then be calculated as 
\begin{equation}
  W^s_{k,k'}(\ve{u})=\sum_{\Delta \ve{n}=-\infty}^{\infty}K^s_{\ve{\Delta n},k,k'}\mathrm{e}^{-2\pi \imu(\ve{u}\cdot\ve{\Delta r_{n}})},
\end{equation}
where \ch{$k$ and $k'$ are indices of two energy bins and} $\ve{\Delta r_{n}}=\ve{r_{n+\Delta n}-r_{n}}$. For $k=k'$, $W^s_{k,k'}(\ve{u})$ is the noise power spectrum (NPS) in energy bin $k$, whereas $W^s_{k,k'}(\ve{u})$ for $k\ne k'$ describes the frequency dependence of correlations between different energy bins.

Now assume that the studied detector is used to discriminate between two cases: a object-absent (background) case with incident photon distribution $q(\ve{r},E)$ and an object-present case where the incident photon distribution is $q^{\mathrm{obj}}(\ve{r},E)=q(\ve{r},E)+\Delta q(\ve{r},E)$, i.e. a signal-known-exactly/background-known-exactly (SKE/BKE) task. The difference in the expected presampled signal between the two cases is $\Delta\overline{d}_{\ve{n},k}$ \ch{with Fourier} transform
\begin{equation}
  \label{eq:signal_difference_as_convolution}
  \Delta \overline{D}_k(\ve{u})=\Delta_x\Delta_y\int_{0}^{E_{\mathrm{max}}}H_k(\ve{u},E) \Delta \overline{Q}(\ve{u},E) \mathrm{d}E,
\end{equation}
where $\Delta \overline{Q}(\ve{u},E)$ is the expected signal difference in the Fourier domain. The expected difference in the sampled signal is $\Delta\overline{d}^s_{\ve{n},k}$ \ch{with discrete-space} Fourier transform $\Delta \overline{D}^s_k(\ve{u})=\frac{1}{\Delta_x\Delta_y}\sum_{m_u,m_v=-\infty}^\infty \Delta \overline{D}_k(u-\frac{m_u}{\Delta_x},v-\frac{m_v}{\Delta_y})$. 

Our measure of the detector performance for this task will be \ch{the} squared signal-difference-to-noise ratio (``detectability'') $d'^2$ of the optimal linear observer, which tells us how well this model observer can discriminate between the two cases (object present or absent) given the available measurements in all pixels and all energy bins. Viewing $\overline{d}^s_{\ve{n},k}$  and $\Delta \overline{d}^s_{\ve{n},k}$ for all $k$ and $\ve{n}$ as elements of vectors $\ve{d}^s$ and $\ve{\Delta \overline{d}}^s$  (see Sec. 13.2.12 of Ref. \citenum{barrett_myers_foundations}), $d'^2$ can be expressed as
\begin{equation}
  \label{eq:d2_spatial_domain}
  d'^2 =\left(\ve{\Delta \overline{d}}^s\right)\trsp \mathrm{Cov}(\ve{d}^s)^{-1} \ve{\Delta \overline{d}}^s.
\end{equation}

\ch{We will now transform (\ref{eq:d2_spatial_domain}) to the Fourier domain and express this formula in a way that separates parameters specific to the detector from parameters specific to the discrimination task. To accomplish this,} we need to assume that signal aliasing is negligible, so that $\Delta \overline{D}^s_k(\ve{u})\approx\frac{1}{\Delta_x\Delta_y}\Delta \overline{D}_k(\ve{u})$. This is true if the pixel size is sufficiently small compared to other resolution-limiting factors, such as the detector point-spread function or the focal spot size, \ch{or if the object is inherently band limited,} or if an oversampling scheme is used i.e. several measurements are performed for different detector positions relative to the object. To obtain an expression that is easily comparable to non-pixelated systems, we also express the noise correlations in terms of \ch{the cross-spectral density $\ma{W}_{d^+}(\ve{u})=\frac{1}{\Delta_x\Delta_y}\ma{W}^s(\ve{u})$ of the sampled pulse train signal $d_k^+(\ve{r})=\sum_{\ve{n}=-\infty}^\infty d^s_{\ve{n},k}\delta(\ve{r}-\ve{r_n})$. These manipulations, which are found in Appendix \ref{sec:appendix_matrix_transformation}, yield}

%Removed, now only found in appendix, 
% \begin{equation}
%     \label{eq:d2_fourier_domain}
%     d'^2= \Delta_x\Delta_y \int_{\mathrm{Nyq}} \sum_k \sum_{k'} \Delta \overline{D}^s_k(\ve{u})^*\left[\left({W^s}\right)^{-1}\right]^*_{k,k'}\!\!(\ve{u}) \Delta \overline{D}^s_{k'}(\ve{u})\mathrm{d}\ve{u},
% \end{equation}

\begin{equation}
  \begin{aligned}
   \label{eq:d2_q_inserted}
    d'^2=  \int_{\mathrm{Nyq}} \sum_k \sum_{k'} \int_0^{E_\mathrm{max}} \int_0^{E_\mathrm{max}} H_k(\ve{u},E)^*\Delta\overline{Q}(\ve{u},E)^*\\
    \cdot\left[W_{d^+}^{-1}\right]^*_{k,k'}\!\!(\ve{u}) H_k(\ve{u},E') \Delta\overline{Q}(\ve{u},E')\mathrm{d}E \mathrm{d}E'\mathrm{d}\ve{u}.
  \end{aligned}
\end{equation}
\ch{where ${}^*$ denotes complex conjugate and $\mathrm{Nyq}$ denotes the Nyquist region $\left\{\ve{u}: \left|u\right|<\frac{1}{2\Delta_x}, \left| v\right|<\frac{1}{2\Delta_y}\right\}$. Note that both $E$ and $E'$ are \textit{incident} energies.}

%Discarded text
% The ideal-linear-observer performance is thus obtained as an integral over spatial frequencies of a quadratic form in $\ve{\Delta \overline{D}}^s=\left(\Delta \overline{D}^s_k(\ve{u})\right)$, the difference vector of registered counts in all energy bins for all spatial frequencies $\ve{u}$. 

To rewrite (\ref{eq:d2_q_inserted}) in a form that is more similar to the corresponding expression for a conventional detector, we introduce the relative signal difference $\Delta S(\ve{u},E) =\Delta\overline{Q}(\ve{u},E)/\overline{q}(E)$ and define a frequency-dependent matrix $\ma{NEQ}(\ve{u})$ with elements given by

\begin{align}
  %\label{eq:neq_definition} %M has beed removed
  %\mathrm{NEQ}(\ve{u},E,E')=\overline{q}(E)M(\ve{u},E,E')\overline{q}(E')\\
  \label{eq:neq_expressed_in_H_and_W}
  \mathrm{NEQ}(\ve{u},E,E')=\sum_k \sum_{k'} \overline{q}(E)H_k(\ve{u},E)^*\left[W_{d^+}^{-1}\right]^*_{k,k'}\!\!(\ve{u}) H_{k'}(\ve{u},E') \overline{q}(E').
\end{align}

This gives

\begin{equation}
   \label{eq:d2_fcn_of_NEQ}
    d'^2= \int_{\mathrm{Nyq}} \int_0^{E_\mathrm{max}} \int_0^{E_\mathrm{max}} \Delta S (\ve{u},E)^*\mathrm{NEQ}(\ve{u},E,E') \Delta S(\ve{u},E')\mathrm{d}E \mathrm{d}E'\mathrm{d}\ve{u},
\end{equation}
which is analogous to the formula for conventional X-ray detectors (Ref. \citenum{cunningham_linear_systems_chapter}, Eq. 2.150, for non-pixelated detectors):
\begin{equation}
   \label{eq:d2_conventional_detector}
    %d'^2= \int_{\mathrm{Nyq}} \left|\Delta S (\ve{u})\right|^2\mathrm{NEQ}(\ve{u})\mathrm{d}\ve{u}
    d'^2= \int \left|\Delta S (\ve{u})\right|^2\mathrm{NEQ}(\ve{u})\mathrm{d}\ve{u},
\end{equation}
%Note on units: Delta S has units mm^2 (relative signal difference per Fourier space area). NEQ has units mm⁻2. The same is true for the vector- and  matrix-valued counterparts in a material basis representation (but not in the energy basis representation).

where (Ref. \citenum{cunningham_linear_systems_chapter}, Eqs. 2.193 and 2.209)
% \begin{equation}
%    \label{eq:neq_conventional_detector}
%     \mathrm{NEQ}= \frac{\overline{q}^2\overline{G}^2 \mathrm{MTF}^2_{\mathrm{pre}}(\ve{u})}{\Delta_x^2\Delta_y^2\mathrm{NPS}_{d^+}(\ve{u})}=\frac{\overline{q}^2 H^2(\ve{u})}{\mathrm{NPS}_{d^+}(\ve{u})}
% \end{equation}
%
\begin{equation}
   \label{eq:neq_conventional_detector}
    \mathrm{NEQ}= \frac{\overline{q}^2\overline{G}^2 \mathrm{MTF}^2_{\mathrm{pre}}(\ve{u})}{\mathrm{NPS}_{\mathrm{dig}}(\ve{u})}=\frac{\overline{q}^2\overline{G}^2 \mathrm{MTF}^2_{\mathrm{pre}}(\ve{u})}{\Delta_x^2\Delta_y^2\mathrm{NPS}_{d^+}(\ve{u})}=\frac{\overline{q}^2 H^2(\ve{u})}{\mathrm{NPS}_{d^+}(\ve{u})}.
\end{equation}

In this equation, $\overline{G}$ is the large-area gain (average registered counts per pixel divided by \ch{the expected input quanta} $\overline{q}$), $\mathrm{MTF}_{\mathrm{pre}}$ is the presampling modulation transfer function and $\left|H(\ve{u})\right|=\overline{G}\,\mathrm{MTF_{pre}}/(\Delta_x\Delta_y)$ is a transfer function defined in analogy with $H_k(\ve{u},E)$.

% %Discarded text: M has beed removed
% By introducing
% \begin{equation}
%   M(\ve{u},E,E')= \sum_k \sum_{k'} H_k(\ve{u},E)^*\left[W_{d^+}^{-1}\right]^*_{k,k'}\!\!(\ve{u}) H_{k'}(\ve{u},E') 
% \end{equation}
% (\ref{eq:d2_q_inserted}) is turned into
% \begin{equation}
%    \label{eq:d2_fcn_of_q}
%     d'^2= \int_{\mathrm{Nyq}} \int_0^{E_\mathrm{max}} \int_0^{E_\mathrm{max}} \Delta\overline{Q}(\ve{u},E)^*\\
%     M(\ve{u},E,E') \Delta\overline{Q}(\ve{u},E')\mathrm{d}E \mathrm{d}E'\mathrm{d}\ve{u}.
% \end{equation}
$\ma{NEQ}(\ve{u})$ is thus a natural generalization of the noise-equivalent quanta used to describe conventional detectors. It is a frequency-dependent matrix with continuous indices $E$ and $E'$ encoding all the information about the detector that is relevant for evaluating the detection performance. It depends on the background case noise characteristics, and thus on the incident spectrum in that setting, but not on the discrimination task. The discrimination task function $\Delta S$ determines how the different frequency components of $\ma{NEQ}$ are weighted together to give the detectability. The $\ma{NEQ}$ matrix therefore defines a quadratic form  in $\Delta S(\ve{u})$ giving the contribution to $d'^2$ at each spatial frequency.

\ch{To attain the performance limit given by $d'^2$, the model observer needs to take data from all the energy bins into account and give an optimal weight to each energy bin at each spatial frequency, which is difficult for a human observer. For any given task, however, frequency-dependent optimal weighting of the energy bin images can be used to form a single image, for which an ideal linear observer attains the same performance for that task.\cite{yveborg_frequency_based_weighting} This observation allows the following interpretation: $d'^2$ is the maximum achievable detectability in a single image that is formed as a weighted sum of the bin images, where the weights are frequency-dependent and optimized for the given detection task. Note that the optimization of the weights ensures that this detectability is always greater than (or equal to) the detectability in an image formed by simply adding the bin images together with equal weight (i.e. an image formed from all counts above the lowest threshold).}

%Removed text - need to go by the spatial domain to avoid questions related to the Fourier transform of a random process.
%$d'^2$ is most conveniently calculated in the Fourier domain by viewing $D^s_k(\ve{u})$  and $\Delta \overline{D}^s_k(\ve{u})$ for all $k$, $u$ and $v$ as elements of vectors $D^s$ and $\Delta \overline{D}^s$ (see Sec. 13.2.12 of \citenum{barrett_myers_foundations}):
%\begin{equation}
%  \begin{aligned}
%    d'^2 =\left(\Delta \overline{D}^s\right)\trsp \mathrm{Cov}(D^s)^{-1} \Delta \overline{D}^s\\
%    =\sum_k \sum_{k'} \iint_{\mathrm{Nyq}}\iint_{\mathrm{Nyq}}\Delta \overline{D}^s_k(\ve{u})
%    \left[\mathrm{Cov}(D^s)^{-1}\right]_{k,k'}\!\!(\ve{u},\ve{u}') \Delta \overline{D}^s_{k'}(\ve{u})\mathrm{d}\ve{u} \mathrm{d}\ve{u}'.
%  \end{aligned}
%\end{equation}
%where $\mathrm{Nyq}$ denotes the Nyquist region $\left\{(\ve{u}): \left|u\right|<\frac{1}{2\Delta_x}, \left| v\right|<\frac{1}{2\Delta_y}\right\}$.

%Discarded text
%Since the noise is assumed to be wide-sence stationary, the covariance matrix $\mathrm{Cov}(D^s)$ has a block-diagonal structure in the sense that different spatial frequencies are uncorrelated: $\left[\mathrm{Cov}(D^s)\right]_{k,k'}\!\!(\ve{u},\ve{u}')=\delta(\ve{u}-\ve{u}')\mathrm{Cov}(D^s_k(\ve{u}),D^s_{k'}(\ve{u}))$

\ch{It is important to keep in mind that both $E$ and $E'$ represent incident energies. It may seem surprising at first that $\mathrm{NEQ}(\ve{u},E,E')$ depends on two incident energy indices instead of one, but this is necessitated by the fact that a detector with imperfect energy resolution may confuse incident photons of different incident energies with each other  (e.g., a photon for which the full energy is recorded and a higher energy photon for which some of the energy escapes). Any description of the detector performance must therefore be able to describe how well the detector can distinguish between incident photons with any combination of energies $E$ and $E'$. The total detectability of a feature is therefore the sum of detectability contributions from all \textit{pairs} of energies $(E,E')$. }

The above description applies to a continuous-to-discrete imaging system, which is the relevant type of system for multibin photon-counting detectors. Note, however, that the analysis, apart from the sampling step, can be adapted to a continuous-to-continuous imaging system by replacing $n_x$, $n_y$ and $k$ by continuous variables. An \textit{ideal} energy-resolving imaging system is a continuous-to-continuous imaging system that registers the correct position and energy for each incoming photon. Its transfer function is therefore $H(\ve{u},E,\varepsilon)=\delta(E-\varepsilon)$, where $\varepsilon$ denotes the registered energy, and its noise cross-spectrum is white, i.e. $W_q(\ve{u},\varepsilon,\varepsilon')$ is constant with respect to $\ve{u}$. For an ideal photon-counting detector, the input signal is Poisson distributed and the registered counts at different energies are uncorrelated, so that $W_q(\ve{u},\varepsilon,\varepsilon')=\overline{q}(\varepsilon)\delta(\varepsilon-\varepsilon')$ where $\overline{q}(\varepsilon)$ is the expected number of incident photons per energy and area (Ref. \citenum{cunningham_linear_systems_chapter}, Sec. 2.6.2.3). For a pixelated but otherwise ideal detector, $W_{d^+}(\ve{u},\varepsilon,\varepsilon')$ tends to $W_q(\ve{u},\varepsilon,\varepsilon')$ as $\Delta_x$ and $\Delta_y$ tend to 0, which gives $\mathrm{NEQ}^{\mathrm{ideal}}(\ve{u},E,E')= \overline{q}(E)\delta(E-E')$ for an ideal detector.
%NOTE: H has a different dimension in the continuous case: 1/energy instead of 1. Similarly, W an extra factor 1/energy^2 in its dimension. Actually, Wd with continuous bins has dimension area/energy^2 whereas W_d+ with discrete bins has dimension 1/area. M has the same dim. as in the finite #bins case (mm2) because of the double integral over E.
%The presampling signal d as we have defined it is undefined for a pixel-free detector, but it turns out that W_d+ converges to W_q=q as the pixel size goes to 0.
% OLD wrong $W^d(\ve{u},E,\varepsilon)=\sigma_q^2\delta(E-\varepsilon)$

In analogy with the conventional definition of DQE, $\mathrm{DQE}(\ve{u})= \mathrm{NEQ}(\ve{u})/\overline{q}$, we can now define a $\ma{DQE}$ matrix by normalizing $\ma{NEQ}$ by the ideal-detector performance. Therefore, let
\begin{equation}
  \label{eq:dqe_definition}
  \mathrm{DQE}(\ve{u},E,E')=\overline{q}(E)^{-1/2}\mathrm{NEQ}(\ve{u},E,E')\overline{q}(E')^{-1/2},
\end{equation}
or equivalently,
\begin{equation}
  \label{eq:dqe_def_expanded}
  \mathrm{DQE}(\ve{u},E,E')= \overline{q}(E)^{1/2}\left[\sum_k \sum_{k'} H_k(\ve{u},E)^*\left[{W^{-1}_{d^+}}\right]^*_{k,k'}\!\!(\ve{u}) H_{k'}(\ve{u},E')\right]\overline{q}(E')^{1/2}
\end{equation}

This definition gives $\mathrm{DQE}^{\mathrm{ideal}}(\ve{u},E,E')=\delta(E-E')$ for the ideal detector. The fact that no detector performs better than an ideal one is now formulated as $\mathrm{\ma{NEQ}} \le \mathrm{\ma{NEQ}^{\mathrm{ideal}}}$ where $\le$ should be interpreted in the matrix inequality sense, i.e. $\ma{A}\le {B}$ if $\ma{B}-\ma{A}$ is positive semidefinite. Proving this inequality mathematically would require analyzing how the noise cross-spectral density is related to the signal transfer function, e.g. using cascaded systems analysis,\cite{cunningham_linear_systems_chapter} which is beyond the scope of this work. We therefore view it as a physically motivated requirement that any realistic detector model must satisfy.
%I have not figured out exactly what the requirements on DQE are. I am not sure that DQE<=I. The diagonal terms are between 0 and 1 of course. The off-diagonal terms must satisfy D21<=sqrt(D11*D12) due to Cauchy-Schwarz but there must be additional constraints to ensure that a change of basis cannot give NEQ>=NEQideal.
%Discarded: and $\mathrm{\ma{DQE}} \le \mathrm{\ma{DQE}}^{\mathrm{ideal}}=\ma{I}$ where $\le$ should be interpreted in the matrix inequality sense, i.e. $\ma{A}\le {B}$ if $\ma{B}-\ma{A}$ is positive semidefinite, and $\ma{I}$ denotes the identity operator.

As a simple example of a nonideal detector, consider a detector that has imperfect quantum detection efficiency $\eta(E)<1$ but infinitely high spatial resolution and perfect energy response. For this detector, $\mathrm{NEQ}(\ve{u},E,E')=\eta(E)\overline{q}(E)\delta(E-E')$ and $\mathrm{DQE}(\ve{u},E,E')=\eta(E)\delta(E-E')$. Our definition of $\ma{DQE}$ can thus be seen as a generalization of the energy-dependent detection efficiency $\eta(E)$.

The diagonal elements of the $\ma{DQE}$ matrix can be interpreted as the dose efficiency per energy for detecting a spectrum change at one single energy, whereas the off-diagonal terms reflect the degree to which different detected energies interfere with each other. This interference may be constructive or destructive, depending on whether the off-diagonal terms are positive or negative. To understand why the $\ma{DQE}$ matrix has dimensions of inverse energy unlike the conventional DQE which is dimensionless, note that a task function which equals $\Delta S(\ve{u})=\Delta S_0$ in a small interval $\Delta E$ around a single energy $E_0$ gives ${d'}^2(\ve{u})\approx\overline{q}(E_0)\mathrm{DQE}(\ve{u},E_0,E_0)\Delta E^2\Delta S_0^2$, which can be compared to $d_\mathrm{ideal}'^2(\ve{u})\approx \overline{q}(E_0)\Delta E \Delta S_0^2$ for the ideal detector. $\mathrm{DQE}^{\mathrm{task}}={d'}^2(\ve{u})/d_\mathrm{ideal}'^2(\ve{u})\approx \mathrm{DQE}(\ve{u},E_0,E_0)\Delta E$ is thus proportional to $\Delta E$ if $\mathrm{DQE}(\ve{u},E_0,E_0)$ is nonsingular. This is the case for binning detectors since these have difficulty detecting a change in a narrow energy band measured in a background of noise contributions from a wide energy interval.

\subsection{NEQ and DQE in material basis}
The $\ma{NEQ}$ matrix contains information about the performance of the detector for any spectral discrimination task. In practice, this matrix would be cumbersome to report, \ch{e.g. in a publication}, especially if the energy scale is discretized into a large number of steps. For example, if the energy variable is discretized in 1 keV-steps from 0 to 140 keV, one needs to supply a $140\times140$ matrix as a function of spatial frequency in order to characterize the detector performance. However, the $\ma{NEQ}$ matrix can be expressed more compactly by exploiting the basis material concept i.e. the approximation that all substances likely to be present in the field of view have attenuation coefficients $\mu(E)$ in a low-dimensional space, spanned by a small number $N_m$ of basis functions $f_l(E)$: $\mu(E)=\sum_{l=1}^{N_m} a_lf_l(E)$.\cite{alvarez_macovski} As long as one is only interested in the detector performance for differentiating materials whose attenuation coefficients lie within this space, and as long as the signal difference is small enough that a linearized description may be used, one only needs to know the restriction of the $\ma{NEQ}$ matrix to this subspace to be able to predict the performance for the relevant tasks.

For a homogeneously illuminated object and in the absence of detected scatter from the object, we can assume that the photon distribution incident on the detector is given by  
\begin{equation}
  \label{eq:forward_model}
  \overline{q}(\ve{r},E)=\Phi(E)\mathrm{exp}\left(-\sum_{l=1}^{N_m}A_l(\ve{r}) f_l(E)\right),
\end{equation}
where $\Phi(E)$ is the number of photons per area and energy incident on the object and $A_l(\ve{r})$ are the basis coefficients integrated along the X-ray beam path. We will study the task of differentiating between two situations, $A^\mathrm{bg}_l(\ve{r})$ and $A^\mathrm{obj}_l(\ve{r})=A^\mathrm{bg}_l(\ve{r})+\Delta A(\ve{r})$, both small perturbations of a homogeneous baseline $A_l(\ve{r})=A_l^0$. In the small-signal approximation, $\overline{q}^{\mathrm{obj}}(\ve{r},E) \approx \overline{q}(E)+\sum_{l=1}^{N_m}\frac{\partial \overline{q}(E)}{\partial A_l} \left(A^{\mathrm{obj}}(\ve{r})-A_l^0\right)$, and similarly for $\overline{q}^{\mathrm{bg}}(\ve{r},E)$. Here, $\overline{q}(E)$ is the expected photon flux at the detector, given by (\ref{eq:forward_model}) for $A_l(\ve{r})=A^0_l$, $l=1,\ldots,N_m$ and the derivative $\frac{\partial \overline{q}(E)}{\partial A_l}$ is evaluated for $A_l=A_l^0$. The difference in expected photon flux between the two cases is approximately $\Delta \overline{q}(\ve{r},E)=\sum_{l=1}^{N_m}\frac{\partial \overline{q}(E)}{\partial A_l}\Delta A_l(\ve{r})$. Letting $\tilde{A}(\ve{u})$ denote the continuous Fourier transform of $A$,
\begin{equation}
\label{eq:fourier_transform_q_and_a_relation}
\Delta \overline{Q}(\ve{u},E)=\sum_{l=1}^{N_m}\frac{\partial \overline{q}(E)}{\partial A_l}\Delta \tilde{A}_l(\ve{u})	
\end{equation}
%Discarded:Let $A_l(\ve{r})= A^0_l+\Delta A_l(\ve{r})$ where $A^1_l(\ve{r})$ is a small perturbation. 

and 
\begin{equation}
\Delta S(\ve{u},E)=\sum_{l=1}^{N_m}L^\mathcal{B}_l(E)\frac{1}{\overline{q}^\mathrm{tot}}\frac{\partial \overline{q}^{\mathrm{tot}}}{\partial A_l}\Delta \tilde{A}_l(\ve{u})=\sum_{l=1}^{N_m}L^\mathcal{B}_l(E)\Delta S^\mathcal{B}_l(\ve{u}),
\end{equation}

where we have defined $L^\mathcal{B}_l(E)=\frac{\overline{q}^\mathrm{tot}}{\overline{q}(E)}\frac{\partial \overline{q}(E)}{\partial A_l}/\frac{\partial \overline{q}^{\mathrm{tot}}}{\partial A_l}$ with $\overline{q}^{\mathrm{tot}}=\int_0^{E_{\mathrm{max}}}\overline{q}(E)\mathrm{d}E$. $L^\mathcal{B}_l(E)$ can be regarded as an element of a transformation matrix from the basis of individual energies to the basis ${\mathcal{B}}$ of differential spectrum changes corresponding to differential path length changes of basis materials $\left\{f_1,f_2,\ldots,f_{N_m}\right\}$. We have also defined the signal difference vector in the basis $\mathcal{B}$, $\ve{\Delta S}^\mathcal{B}(\ve{u})$, by $\Delta S_l^{\mathcal{B}}(\ve{u})=\frac{1}{\overline{q}^\mathrm{tot}}\frac{\partial \overline{q}^{\mathrm{tot}}}{\partial A_l}\Delta \tilde{A}_l(\ve{u})$. Eq. \ref{eq:d2_fcn_of_NEQ} can now be expressed as 

%Old definition, discarded
%$\tilde{\ve{A}}^\mathrm{norm}(\ve{u})=\left(\frac{\Delta \tilde{A}_1(\ve{u})}{A_1^0},\ldots,\frac{\Delta \tilde{A}_{N_m}(\ve{u})}{A_{N_m}^0}\right)\trsp$, and $\ve{\Delta \tilde{A}}(\ve{u})$ as the corresponding difference vector. Using (\ref{eq:d2_fcn_of_NEQ}), we now obtain

\begin{align}
   \label{eq:d2_fcn_of_NEQB}
    d'^2=\frac{1}{{\overline{q}^\mathrm{tot}}^2} \int_{\mathrm{Nyq}} \sum_l \sum_{l'} \Delta \tilde{A}_l(\ve{u})^* \frac{\partial \overline{q}^{\mathrm{tot}}}{\partial A_l} \mathrm{NEQ}_{l,l'}^{\mathcal{B}}(\ve{u}) \frac{\partial \overline{q}^{\mathrm{tot}}}{\partial A_{l'}}\Delta \tilde{A}_{l'}(\ve{u}) \mathrm{d}\ve{u} \\ \nonumber
    =\int_{\mathrm{Nyq}} \ve{\Delta S}^\mathcal{B}(\ve{u})^\dagger \ma{NEQ}^\mathcal{B}(\ve{u})\ve{\Delta S}^\mathcal{B}(\ve{u})\mathrm{d}\ve{u},
\end{align}
%Removed star on  q derivative since this is a real quantity

where the matrix elements of the $\ma{NEQ}$ matrix in basis ${\mathcal{B}}$ are given by
\begin{equation}
  \mathrm{NEQ}_{l,l'}^{\mathcal{B}}(\ve{u})=\int_0^{E_\mathrm{max}} \int_0^{E_\mathrm{max}} L^\mathcal{B}_l(E) \mathrm{NEQ}(\ve{u},E,E') L^\mathcal{B}_{l'}(E') \mathrm{d}E \mathrm{d}E',
\end{equation}
or, in a form that is easier to compute:
\begin{equation}
    \label{eq:NEQ_in_basis}
    \mathrm{NEQ}_{l,l'}^{\mathcal{B}}(\ve{u})= \sum_k \sum_{k'} H^\mathcal{B}_{k,l}(\ve{u})^*\left[W_{d^+}^{-1}\right]^*_{k,k'}\!\!(\ve{u}) H^\mathcal{B}_{k',l'}(\ve{u}).
\end{equation}
Here, $H_{k,l}^\mathcal{B}(\ve{u}) = \int_{E=0}^{E_{\mathrm{max}}} \overline{q}^\mathrm{tot}\left(\frac{\partial \overline{q}(E)}{\partial A_l}/\frac{\partial \overline{q}_{\mathrm{tot}}}{\partial A_l} \right)H_k(\ve{u},E) \mathrm{d}E$ is the transfer function, relating the magnitude of a small relative modulation of the input photon fluence corresponding to a change in $A_l$ at spatial frequency $\ve{u}$ to the resulting change in registered counts per area in energy bin $k$ at that frequency.

Like the $\mathrm{NEQ}$ for non-energy resolving detectors, $\ma{NEQ}^{\mathcal{B}}$ has units of area$^{-1}$. Each diagonal component $\mathrm{NEQ}^\mathcal{B}_{ll}$ of the latter specifies the number of quanta per area that an ideal detector needs to measure to achieve the same detectability as the studied detector, when the task is to detect the addition of a small amount of basis material $l$.
% Discarded: the spectrum change caused by (the addition of a small amount...)
The $\ma{DQE}$ matrix in basis $\mathcal{B}$ is defined in analogy with (\ref{eq:dqe_definition}):
\begin{equation}
  \label{eq:DQE_in_basis}
  \mathrm{DQE}^{\mathcal{B}}_{l,l'}(\ve{u})=\frac{\mathrm{NEQ}^{\mathcal{B}}_{l,l'}(\ve{u})}{\sqrt{\mathrm{NEQ}^{\mathcal{B},\mathrm{ideal}}_{l,l} \cdot \mathrm{NEQ}^{\mathcal{B},\mathrm{ideal}}_{l'l'}}},
\end{equation}
%Obsolete definition:
% \begin{equation}
%   \ma{DQE}^{\mathcal{B}}(\ve{u})=\left[\ma{NEQ}^{\mathcal{B},\mathrm{ideal}}\right]^{-1/2}\ma{NEQ}^{\mathcal{B}}(\ve{u})\left[\ma{NEQ}^{\mathcal{B},\mathrm{ideal}}\right]^{-1/2}
% \end{equation}
where our notation reflects that $\ma{NEQ}^{\mathcal{B},\mathrm{ideal}}$ is independent of spatial frequency.
%Half-finished eq. Do not use!
%\begin{equation}
%   \label{eq:fisher}
%    F_{lm}(\ve{u})=\sum_k \sum_{k'}\frac{\partial \left[\left({W_{\hat{d}}}\right)^{-1}\right]_{k,k'} \frac{\partial \overline{D}_k^s(\ve{u})}{A(\ve{u})}}+\frac{1}{2}\sum_{k}
%\end{equation}
%Are the below comments old?
The $\ma{DQE}$ defined in this way is a dose-normalized quantity that encodes the performance of the detector for all different tasks. For any specific detection task, the performance relative to an ideal detector is given by the task-specific DQE, which is a scalar-valued quantity:
\begin{align}
  \label{eq:dqe_task_detection}
  \mathrm{DQE}^{\mathrm{task}}(\ve{u}) = \frac{d'^2(\ve{u})}{d_\mathrm{ideal}'^2(\ve{u})}=\frac{\int_0^{E_\mathrm{max}} \int_0^{E_\mathrm{max}} \Delta S (\ve{u},E)^*\mathrm{NEQ}(\ve{u},E,E') \Delta S(\ve{u},E')\mathrm{d}E \mathrm{d}E'}{\int_0^{E_\mathrm{max}} \left|\Delta S (\ve{u},E)\right|^2\overline{q}(E) \mathrm{d}E} \\ \nonumber
  = \frac{\ve{\Delta S}^\mathcal{B}(\ve{u})^\dagger \ma{NEQ}^{\mathcal{B}}(\ve{u}) \ve{\Delta S}^\mathcal{B}(\ve{u})}{\ve{\Delta S}^\mathcal{B} (\ve{u})^\dagger \ma{NEQ}^{\mathcal{B},\mathrm{ideal}}(\ve{u}) \ve{\Delta S}^\mathcal{B}(\ve{u})},
  %OLD: = \frac{\ve{\Delta A}^{\mathrm{norm}} (\ve{u})^\dagger \ma{NEQ}^{\mathcal{B}}(\ve{u}) \ve{\Delta A}^\mathrm{norm}(\ve{u})}{\ve{\Delta A}^{\mathrm{norm}} (\ve{u})^\dagger \ma{NEQ}^{\mathcal{B},\mathrm{ideal}}(\ve{u}) \ve{\Delta A}^{\mathrm{norm}}(\ve{u})}
  %= \frac{\sum_l \sum_{l'}\frac{\Delta A_l (\ve{u})^*}{A_{l}^0}\mathrm{NEQ}_{l,l'}^{\mathcal{B}}(\ve{u}) \frac{\Delta A_{l'}(\ve{u})}{A_{l'}^0}}{\sum_l \sum_{l'} \frac{\Delta A_l (\ve{u})^*}{A_{l}^0}\mathrm{NEQ}_{l,l'}^{\mathcal{B},\mathrm{ideal}}(\ve{u}) \frac{\Delta A_{l'}(\ve{u})}{A_{l'}^0}}
\end{align}
where we have used the notation $d'^2(\ve{u})$ for the $d'^2$ contribution from frequency $\ve{u}$. Eq. \ref{eq:dqe_task_detection} shows that $\mathrm{DQE}^{\mathrm{task}}$ for detecting a small difference at frequency $\ve{u}$ in one of the basis materials used in the decomposition,  basis material $l$, is simply given by the diagonal element $\mathrm{DQE}^{\mathcal{B}}_{l,l}(\ve{u})$ of $\ma{DQE}^\mathcal{B}$.
\ch{The off-diagonal terms of $\mathrm{DQE}^{\mathcal{B}}_{l,l}$, on the other hand, specify the degree of constructive or destructive interference when the task involves detecting a difference in two or more basis functions. A positive off-diagonal term corresponds to a positive interference effect on detectability when both basis material path lengths are increased simultaneously.}

%Old text, moved: For a given basis $\mathcal{B}$, the diagonal elements of $\ma{DQE}^\mathcal{B}$ tell us how efficient the detector is for detecting a change in only one of the basis functions, 

$\mathrm{DQE}^{\mathrm{task}}$ for detecting differences due to a material with another spectral response can be obtained by expressing the DQE matrix in any basis which includes the linear attenuation of that feature as a basis function. When transforming to another basis of differential spectrum changes, $\ma{NEQ}$ transforms according to the normal rules for coordinate changes of quadratic forms, but there is no simple transformation rule for $\ma{DQE}$ since its components are obtained as ratios of components of two matrices both of which are basis-dependent.

%OLD, obsolete text
% Unfortunatly, there is in general no simple relation between the matrix-valued DQE and $\mathrm{DQE}^{\mathrm{task}}$, although the latter can be computed from the former with the matrix-valued $\mathrm{NEQ}$ as an intermediate step. We are therefore led to conlude that the matrix-values $\mathrm{NEQ}$ is a more useful quantity than the matrix-values $\mathrm{DQE}$.

\ch{In particular}, we note that Eq. \ref{eq:dqe_task_detection} can be used to calculate the maximum and minimum $\mathrm{DQE}^{\mathrm{task}}$ for any detection task \ch{where the linear attenuation coefficient of the feature to detect is a linear combination of the basis functions in $\mathcal{B}$}. The change of variables $\ve{\Delta S}^\mathcal{B'} (\ve{u})= \left[\ma{NEQ}^{\mathcal{B},\mathrm{ideal}}(\ve{u})\right]^{1/2} \ve{\Delta S}^\mathcal{B} (\ve{u})$ transforms (\ref{eq:dqe_task_detection}) into 
\begin{equation}
\mathrm{DQE}^{\mathrm{task}}(\ve{u}) = \left[\ve{\Delta S}^\mathcal{B'}(\ve{u})^\dagger {\ma{NEQ}^{\mathcal{B},\mathrm{ideal}}(\ve{u})^\dagger}^{-1/2} \ma{NEQ}^{\mathcal{B}}(\ve{u}) \ma{NEQ}^{\mathcal{B},\mathrm{ideal}}(\ve{u})^{-1/2} \ve{\Delta S}^\mathcal{B'}(\ve{u})\right]/\left\Vert\ve{\Delta S}^\mathcal{B'} \right\Vert^2,
\end{equation}
with maximum and minimum values given by the eigenvalues of 

${\ma{NEQ}^{\mathcal{B},\mathrm{ideal}}(\ve{u})^\dagger}^{-1/2} \ma{NEQ}^{\mathcal{B}}(\ve{u}) \ma{NEQ}^{\mathcal{B},\mathrm{ideal}}(\ve{u})^{-1/2}$.

\subsection{Material decomposition}
We now leave the study of detection-task performance for a moment and turn to another question: how well can the material composition of an object be quantified using measurements with the detector? One could try to estimate $A_l(\ve{r})$ directly, but a more fruitful approach is to estimate its Fourier transform $\tilde{A}_l(\ve{u})$. A limit on estimation performance is given by the Cram\'er-Rao lower bound (CRLB), which gives a lower limit of the covariance matrix of the estimated parameter for any unbiased estimator and has been shown to be a useful tool for analyzing estimation accuracy for spectral CT tasks\cite{roessl_hermann_crlb}. Although the CRLB only provides a lower bound, in practice the variance of the maximum likelihood estimator usually agrees well with the CRLB for material decomposition tasks, as has been shown in simulations\cite{roessl_sensitivity_of_k_edge}. The authors are not aware of a tractable form for the CRLB for correlated Poisson variables in general, but as long as there are at least a few tens of photons in every energy bin and measurement, one can approximate the joint probability distribution of the bin counts as multivariate Gaussian. This approximation is valid both with and without pileup taken into account, but in the pileup-free case both mean and variance of each individual measurement will be equal to $d^s_{\ve{n},k}$. The components of the Fisher information matrix $\ma{F_{\bm{\uptheta}}}$ for a real-valued vector parameter $\bm{\uptheta}$ from a measured multivariate Gaussian random variable with mean $\bm{\upmu}$ and covariance matrix $\ma{\Sigma}$ is given by: \cite{kay_estimation_theory}

\begin{equation}
  \label{eq:CRLB_gaussian_full}
  \left[\ma{F_{\bm{\uptheta}}}\right]_{ij} = \left(\frac{\partial \bm{\upmu}}{\partial \uptheta_i}\right)\trsp \ma{\Sigma}^{-1}\frac{\partial \bm{\upmu}}{\partial \uptheta_j} + \mathrm{Tr}\left( \ma{\Sigma}^{-1}\frac{\partial \ma{\Sigma}}{\partial \uptheta_i} \ma{\Sigma}^{-1} \frac{\partial \ma{\Sigma}}{\partial \uptheta_j} \right).
\end{equation}

The CRLB now states that $\mathrm{Cov}(\bm{\hat{\uptheta}}) \ge \ma{F_{\bm{\uptheta}}}^{-1}$ for any unbiased estimator $\bm{\hat{\uptheta}}$ of $\bm{\uptheta}$.
%Discarded $\bm{\uptheta}=\left(\bm{\uptheta}_1, \bm{\uptheta}_2\right)\trsp$ to be estimated contains the real part ($\bm{\uptheta}_1=\mathrm{Re}\;\tilde{\ve{A}})$ and imaginary part ($\bm{\uptheta}_2=\mathrm{Im}\;\tilde{\ve{A}})$) of the vector $\tilde{\ve{A}}$

In our case, $\bm{\upmu}$ and $\ma{\Sigma}$ are the mean and covariance of the measured counts vector $\ve{d}^s$, while the parameter $\bm{\uptheta}=\left(\mathrm{Re}\;\tilde{\ve{A}}, \mathrm{Im}\;\tilde{\ve{A}}\right)\trsp$ to be estimated contains the real and imaginary parts of the vector $\tilde{\ve{A}}$ obtained by concatenating the vectors $\tilde{\ve{A}}(\ve{u})$ for all spatial frequencies in the intersection of the right half-plane $\left\{\ve{u}: u > 0 \;\mathrm{or}\; u=0,v>0\right\}$ and the Nyquist region. Since $\tilde{\ve{A}}$ is the Fourier transform of a real-valued function and hence conjugate-symmetric in $\ve{u}$, we only estimate it on half the Nyquist region in order to avoid getting a singular covariance matrix. \ch{(Readers who are skeptical about our use of the CRLB derived in a finite-dimensional setting for estimating a function in an infinite-dimensional space can think of the function $\tilde{\ve{A}}(\ve{u})$ as being approximated by a large but finite number of delta functions at discrete sample points in the $\ve{u}$ plane. $\tilde{\ve{A}}$ then becomes a vector of delta function coefficients.)}

In the absence of pileup, we note that both $\ve{\overline{d}}^s$ and $\mathrm{Cov}({\ve{d}^s})$, and therefore also the first term in (\ref{eq:CRLB_gaussian_full}), are proportional to the pre-patient photon flux, whereas the second term is independent of this quantity. As long as the photon fluence is high enough, we can therefore approximate the second term with zero. We then obtain the Fisher information matrix $\ma{F}_{\bm{\uptheta}}$ for $\bm{\uptheta}$ as

\begin{equation}
  \label{eq:fisher_first_term}
  \ma{F_{\bm{\uptheta}}} \approx \left(\frac{\partial \ve{\overline{d}}^s}{\partial \bm{\uptheta}}\right)\trsp \mathrm{Cov}(\ve{d}^s)^{-1}\frac{\partial \ve{\overline{d}}^s}{\partial \bm{\uptheta}}.
\end{equation}

\ch{As shown in Appendix \ref{sec:appendix_fisher_matrix}, this leads to the following expression for the CRLB for the components of $\tilde{\ve{A}}(\ve{u})$:}
\begin{equation}
  \label{eq:CRLB_and_NEQ}
  \mathrm{Cov}\left(\hat{\tilde{A}}_l(\ve{u}),\hat{\tilde{A}}_{l'}(\ve{u'})\right) \ge {{\overline{q}^\mathrm{tot}}^2}\frac{\left[\ma{NEQ}^{\mathcal{B}}(\ve{u})^{-1}\right]_{l,l'}}{ \frac{\partial\overline{q}^{\mathrm{tot}}}{\partial A_l}\frac{\partial\overline{q}^{\mathrm{tot}}}{\partial A_{l'}}}\delta(\ve{u}-\ve{u'}).
  %Obsolete: \mathrm{Cov}\left(\hat{\tilde{\ve{A}}}^{\mathrm{norm}}(\ve{u})\right) \ge \left[\ma{NEQ}^{\mathcal{B}}(\ve{u})\right]^{-1}
\end{equation}
\ch{Here, the complex covariance is defined as $\mathrm{Cov}(z,w)=\mathrm{E}\left[(z-\mathrm{E}(z))(w-\mathrm{E}(w))^*\right]$ where $\mathrm{E}$ denotes expected value.}
%NOTE: This should be the only place where the definition of complex covariance actually matters. This definition is consistent with Schreier and Scharf eq 3.3.
\chtwo{Eq. \ref{eq:CRLB_and_NEQ} specifies the achievable variance and covariance in the basis images, and thereby answers our question about the material quantification performance. It is a natural extension of the commonly used zero-frequency CRLB\cite{roessl_hermann_crlb} to frequency-dependent estimation tasks, which allows us to describe the full noise correlation structure in the material-decomposed images, rather than just the pixel-wise variance.}

In analogy with the definition (\ref{eq:dqe_task_detection}) of $\mathrm{DQE}^{\mathrm{task}}$ for a \textit{detection} task we can define the $\mathrm{DQE}^{\mathrm{task}}$ for \textit{quantification} of basis material $l$ as 
\begin{equation}
  \label{eq:dqe_task_quantification}
  \mathrm{DQE}^{\mathrm{task}}(\ve{u})=\frac{\mathrm{Var}(\hat{\tilde{A}}_l^\mathrm{ideal})}{\mathrm{Var}(\hat{\tilde{A}}_l(\ve{u}))},
\end{equation}
where $\mathrm{Var}\left(\hat{\tilde{A}}_l^\mathrm{ideal}\right)$ is the CRLB of the variance of $\tilde{A}_l$ when measured with an ideal detector.

\ch{Eq. \ref{eq:CRLB_and_NEQ} shows that $\ma{NEQ}^{\mathcal{B}}(\ve{u})$, which describes performance for detection tasks, and the frequency-dependent CRLB, which describes the detector performance for material quantification tasks, are very closely related. We will now conclude our theory presentation with a demonstration of another close connection between the detection and material quantification frameworks. To this end, we note that it is possible to make a frequency-dependent optimally weighted sum of the decomposed basis images, just as this can be done with the original energy bin images, and we then investigate what detection performance can be achieved in the weighted image if the weights are chosen optimally.} The lower bound \ch{for the variance} in the basis images given by the CRLB translates to an upper bound of the achievable detectability in any such weighted basis image. \ch{This upper bound is given by the optimal-linear-observer detectability\cite{barrett_myers_foundations} and can be expressed as (see Appendix \ref{sec:appendix_CRLB_material_images}):}
\begin{align}
  \label{eq:d2_upper_limit_weighted_basis_image}
  d'^2 \le \int_\Rtwo \ve{\Delta S}^{\mathcal{B}}(\ve{u})^\dagger \ma{NEQ}^{\mathcal{B}}(\ve{u}) \ve{\Delta S}^{\mathcal{B}}(\ve{u}) \mathrm{d}\ve{u}.
\end{align}

By comparing (\ref{eq:d2_upper_limit_weighted_basis_image}) with (\ref{eq:d2_fcn_of_NEQB}), we see that the same detectability is obtained by combining the different basis images as from combining the original energy bin images, as long enough statistics is available that the estimator variance will be close to the CRLB. This is a generalization of the analogous result for zero frequency derived by Alvarez.\cite{alvarez_near_optimal} \ch{We therefore conclude that the process of basis material decomposition preserves the information contained in the image, as long as the variance of the estimator is close to attaining the CRLB.}

\ch{We emphasize that while this optimal weighting of basis images is theoretically interesting, in many situations there are other ways of using the basis images that are preferable. For example, the estimated basis images can be viewed directly as they are, e.g. as contrast agent distribution maps or virtual non-contrast (VNC) images. Although each individual basis image has lower detectability for specific tasks compared to a frequency-dependent weighted sum of all basis images, the individual images may be more useful in practice since they provide information about material composition \ch{and may provide higher lesion conspicuity}. Another way to use the basis images is to generate virtual monoenergetic images (VMI)}\cite{alvarez_seppi, lehmann_alvarez} as $p^{\mathrm{mono}}(E) =\sum_{l=1}^{N_m}A_l(\ve{r}) f_l(E)$. This is useful because the image values are easy to interpret. Note, however, that the basis images in this case are combined using frequency-independent weight functions, meaning that there is no guarantee that there is an energy for which the resulting monoenergetic image has optimal detectability.
%NOTE I used p mono for line integral of mu, in analogy with the use fo p for log normalized counts.
%Old, can be removed
%As commonly done in basis material decomposition, we assume that the incoming photon distribution has been filtered through a combination of a small number $N_m$ of basis materials: $q(\ve{r},E)=\int_0^{E_\mathrm{max}}\Phi(E)\mathrm{exp}\left(-\sum_{l=1}^{N_m}A_l(\ve{r}) f_l(E)\right)\mathrm{d}E$. In the small-signal approximation, $Q(\ve{u},E)=Q^0(\ve{u},E)+\sum_{l=1}^{N_m}\frac{\partial Q(E)}{\partial A_l}\Delta A_l(\ve{u})$ where the derivative $\frac{\partial Q(E)}{\partial A_l}$ is independent of $u$ and $v$. Since spatial frequencies can be treated independently, the Fisher information matrix for estimating $A_l(\ve{u})$ from the Fourier transformed measured data $D^s_k(\ve{u})$ is derived in analogy with \cite{roessl_hermann_crlb}:

\section{Materials and Methods}
\label{sec:materials_and_methods}
To illustrate the matrix-valued NEQ and DQE measures, we will study their form for two simple models (one modeling K-escape and one modeling Compton scatter in the detector), and also for a more realistic simulation model of a CdTe detector. \ch{Each of these three models is described in one subsection (\ref{sec:methods_fluorescence_model}-\ref{sec:methods_cdte_simulation}) below, and the results for each model are presented and discussed in corresponding subsections of Sec. \ref{sec:results} and \ref{sec:discussion}.}
\subsection{Fluorescence model}
\label{sec:methods_fluorescence_model}
As a simplified model of fluorescence K-escape, we assume that an incident photon of energy $E$ will deposit either its entire energy, with probability $1-p_F(E)$, or an energy $E-E_F$ with probability $p_F(E)$. $E_F$ is thus the energy of the fluorescence photon, and the probability of fluorescence escape $p_F(E)$ is a function of $E$ with $p_F(E)=0$ for $E \le E_F$. For simplicity, we neglect detection of secondary (fluorescence) photons in other detector elements, \ch{e.g. all fluorescence photons escape or} the detector is assumed to have coincidence logic that eliminates secondary events but is not able to reconstruct the original photon energies. We also assume that the detector has infinitely many bins, infinitely small pixels and perfect energy resolution, i.e. the fluorescence escape is the only degrading factor. The transfer function is then $H(\ve{u},E,\varepsilon)=(1-p_F(E))\delta(E-\varepsilon)+p_F(E)\delta(E-E_F-\varepsilon)$ and an incident spectrum $\overline{q}(E)$ results in a measured spectrum $\overline{d}(\varepsilon)=(1-p(\varepsilon))\overline{q}(\varepsilon)+p(\varepsilon+E_F)\overline{q}(\varepsilon+E_F)$. Since each photon is only registered once, the measurements at different energies are independent random variables, and the cross-spectral density is given by $W(\ve{u},\varepsilon,\varepsilon')=\overline{d}(\varepsilon)\delta(\varepsilon-\varepsilon')$. Inserting these results into (\ref{eq:neq_expressed_in_H_and_W}) gives, after some algebra,

\begin{align}
  \label{eq:fluorescence_model}
  \mathrm{NEQ}(\ve{u},E,E')=\left[ \frac{(1-p_F(E))^2}{\overline{d}(E)}+\frac{p_F(E)^2}{\overline{d}(E-E_F)}\right]\overline{q}(E)^2\delta(E-E')+ \nonumber\\
  \frac{(1-p_F(E))\overline{q}(E)p_F(E')\overline{q}(E')}{\overline{d}(E)}\delta(E'-(E+E_F))+ \nonumber\\ \frac{p_F(E)\overline{q}(E)(1-p_F(E'))\overline{q}(E')}{\overline{d}(E')}\delta(E'-(E-E_F)).
\end{align}
The DQE can then be obtained from (\ref{eq:dqe_definition}).

To plot the $\ma{DQE}$ matrix for a simple example, we discretized energy in steps of 1 keV and assumed $E_F=25 \; \mathrm{keV}$ and $p_F(E)=0.2$ (independent of $E$) for $E > E_F$. For simplicity we also assumed a rectangular incident spectrum: $q(E)=\overline{q}^{0}=\overline{q}^{\mathrm{tot}}/(E_2-E_1)$ for $E_1 < E < E_2$ and 0 otherwise. In order to study the effect of spectral overlap, we studied two different incident spectra: one that is nonzero between $E_1=40 \; \mathrm{keV}$ and $E_2=60 \; \mathrm{keV}$, which gives nonoverlapping deposited spectra for the K-escape peak and photopeak events, and one that is nonzero between $E_1=40 \; \mathrm{keV}$ and $E_2=120 \; \mathrm{keV}$, which produces overlap between the photopeak and K-escape spectra. We also used (\ref{eq:dqe_task_detection}) to calculate $\mathrm{DQE^{task}}$ for different task functions $\Delta S(E)$ which were, for simplicity, taken to be piecewise constant, equal to either 0 or $\pm \Delta S_0$, as described below.
\subsection{Scatter model}
\label{sec:methods_scatter_model}
In order to demonstrate how the proposed framework treats interactions where the energy information is lost, we also studied a simple model for Compton scatter in the detector. In this model, we assume that an incident photon is either photoabsorbed with probability $1-p_S$, depositing all its energy, or Compton scattered \ch{in the detector} with probability $p_S=0.5$. In the latter case it deposits an energy $E_S$ which we for simplicity assume to be fixed at $10 \; \mathrm{keV}$ independent of the incident energy. We also assume that all scattered photons are stopped by blocking lamellae or escape from the detector, so that each photon is counted only once. Like in the fluorescence model, we assumed that the incident spectrum is constant, equal to $\overline{q}^{0}=\overline{q}^{\mathrm{tot}}/(E_2-E_1)$ between $E_1=40 \; \mathrm{keV}$ and $E_2=120 \; \mathrm{keV}$ and 0 otherwise. Once again, the photopeak and distorted spectra do not overlap. \ch{This gives the detected spectrum as} $\overline{d}(\varepsilon)=p_S \overline{q}^{\mathrm{tot}}\delta(E-E_s)+(1-p_S)\overline{q}(E)$ and, after some calculations, $\mathrm{DQE}(E,E')=\frac{p_S}{E_1-E_2}+(1-p_S)\delta(E-E')$.

\subsection{CdTe simulation}
\label{sec:methods_cdte_simulation}
To study a more realistic case, we used Monte Carlo simulation (pyPENELOPE\cite{pypenelope}, based on PENELOPE\cite{penelope}) to model a CdTe detector. \ch{This simulation model is similar to the model studied previously in Ref. \citenum{rajbhandary_frequency_dependent_dqe}.} A special version of pyPENELOPE was compiled in order to include secondary photons. For each 1 keV-step from 1 to 120 keV, a pencil beam of photons incident on a 3 mm thick slab of CdTe was simulated, and the location and energy of each photon interaction \ch{were recorded.}
% Removed, confusing: Care was taken to avoid double-counting energy released as K-escape photons.
%Discarded: 10000 particle trajectories were simulated, and for the trajectories belonging to photons among these (decreasing from 9986 trajectories for 1 keV to 4212 for 120 keV) 

\ch{We} simulated charge sharing according to the uniform spherical charge cloud model of Taguchi\cite{taguchi_spatioenergetic}, where the diameter $d$ of the charge cloud is related to the deposited energy $E$ as $d=d_0(E/E_{\mathrm{ref}})^{1/3}$, and we used $d_0=30 \; \upmu \mathrm{m}$ at $E_{\mathrm{ref}} = 70 \; \mathrm{keV}$. By dividing the detector into $0.5\times0.5 \; \mathrm{mm}^2$ pixels and calculating the charge cloud volume fraction located within the borders of each pixel, we obtained the total charge collected in each pixel. The charge contributions from all interactions stemming from each individual photon were then summed, and the total number of registered photons in each of five energy bins (25-40, 41-54, 55-64, 65-77 and 78-120 keV) were recorded. We repeated this for a grid of $16\times16$ positions of the pencil beam in one quadrant of the detector cell, extending from the center to the pixel border in the positive $x$ and $y$ directions. Exploiting mirror symmetry in two dimensions, we obtained the point-spread function $h_k(\ve{\Delta r},E)$ on a grid with $30\times30$ sample points for each detector cell in a grid of $3\times3$ cells. The point-spread function was also symmetrized with respect to interchanging the $x$ and $y$ coordinates.

To obtain the autocovariance function, we used the same pencil beam simulation but with a randomized position of each incident photon to simulate homogeneous illumination of the center pixel. For each incident energy $E$, we recorded the number of incident photons  $N_{\ve{n},k,\ve{n'},k'}(E)$ leading to a registered count in both energy bin $k$ of pixel $\ve{n}$ and energy bin $k'$ of pixel $\ve{n'}=\ve{n}+\ve{\Delta n}$. The covariance of the number of registered photons $N_{\ve{n},k}(E)$ and $N_{\ve{n'},k'}(E)$ in these two pixel-bin pairs is given by $\mathrm{Cov}\left({N_{\ve{n},k}(E),N_{\ve{n'},k'}(E)}\right)=N_{\ve{n},k,\ve{n'},k'}(E)$, since the photons registered in either of $N_{\ve{n},k}(E)$ and $N_{\ve{n'},k'}(E)$, but in not both, are independent between these two measurements. By summing over $\ve{n}$ in the grid of $3\times3$ detector cells, we simulated homogeneous illumination of the entire pixel array and thereby obtained the autocovariance function $K^s_{\ve{\Delta n},k,k'}(E)$ for each incident energy $E$. $K^s_{\ve{\Delta n},k,k'}(E)$ was calculated for $-1\le n_x, n_y \le 1$, i.e. only correlations between nearest neighbors was taken into account. We symmetrized the autocovariance function using mirror symmetry in $x$ and $y$ and with respect to interchanging the $x$ and $y$ coordinates.

%For each pair of energy bins and pixels $(\ve{n},k)$ and $(\ve{n+\Delta n},k')$, we also stored the number of incident photons registered in both. This allowed us to obtain the autocovariance function, $K^s_{\ve{\Delta n},k,k'}$ for each incident monoenergy, exploiting the fact that the covariance between any pair of energy bins in any pair of pixels is equal to the number of expected photons registered in both, for a multivariate Poisson distribution.

We used discrete Fourier transformation to calculate the transfer function $H_k(\ve{u},E)$ and energy-dependent cross-spectral density $W^s_{k,k'}( \ve{u},E)$. We then calculated the noise cross-spectral density of the bin counts $W^s_{k,k'}( \ve{u})$ by combining contributions from different energies using a 120 kVp tungsten anode X-ray spectrum with a prepatient photon fluence of $4\cdot10^6 \; \mathrm{mm}^{-2}$ and $12^\circ$ anode angle, filtered through 2.5 mm Al (prepatient filtration) and 100 mm water.\cite{hernandez_boone_tasmics} The spectrum was obtained from Spektr 3.0\cite{spektr3} and the linear attenuation coefficients were obtained from NIST\cite{nist_xcom}. Electronic noise and pileup were not included in the simulation. The $\ma{NEQ}$ and $\ma{DQE}$ matrices were calculated in the basis of monoenergies using Eqs. \ref{eq:neq_expressed_in_H_and_W} and \ref{eq:dqe_definition} and in the basis $\mathcal{B}=\left\{\mathrm{Water},\mathrm{Bone}\right\}$ using Eqs. \ref{eq:NEQ_in_basis} and \ref{eq:DQE_in_basis}. By calculating the eigenvalues of ${\ma{NEQ}^{\mathcal{B},\mathrm{ideal}}(\ve{u})^\dagger}^{-1/2} \ma{NEQ}^{\mathcal{B}}(\ve{u}) \ma{NEQ}^{\mathcal{B},\mathrm{ideal}}(\ve{u})^{-1/2}$, we obtained the the maximum and minimum $\mathrm{DQE}^\mathrm{task}$ for any detection task. We also calculated the task-specific DQEs for quantifying the amount of cortical bone and water in a two-basis decomposition using Eqs. \ref{eq:CRLB_and_NEQ} and \ref{eq:dqe_task_quantification}.

\section{Results}
\ch{In the below sections \ref{sec:results_fluorescence_model}-\ref{sec:results_cdte_simulation}, we present the results of the calculations of NEQ and DQE for each of the three models described in Sec. \ref{sec:methods_fluorescence_model}-\ref{sec:methods_cdte_simulation}: the fluorescence and scatter toy models and the CdTe model.}
\label{sec:results}
\subsection{Fluorescence model}
\label{sec:results_fluorescence_model}
The results for the fluorescence model are shown in Fig. \ref{fig:flourToyModel}, \ch{with} Fig. \ref{fig:flourToyModel}(a-c) for the nonoverlapping spectrum case \ch{and Fig.} \ref{fig:flourToyModel}(d-i) for the overlapping spectrum case. In Fig. \ref{fig:flourToyModel}(c,f-i), different task functions are shown together with the diagonal of the DQE matrix and the corresponding $\mathrm{DQE^{task}}$ values. The $\ma{DQE}$ matrix was discretized by replacing the delta function with a square of width $1\; \mathrm{keV}$. For the overlapping spectrum case (Fig. \ref{fig:flourToyModel}e), the delta function coefficients in the $\mathrm{DQE}$ matrix are $0.68 \; - \; 0.84$ on the diagonal and $0.16$ on the off-diagonal lines.

\subsection{Scatter model}
\label{sec:results_scatter_model}
\begin{figure}
  \centering
  \includegraphics[scale=0.60]{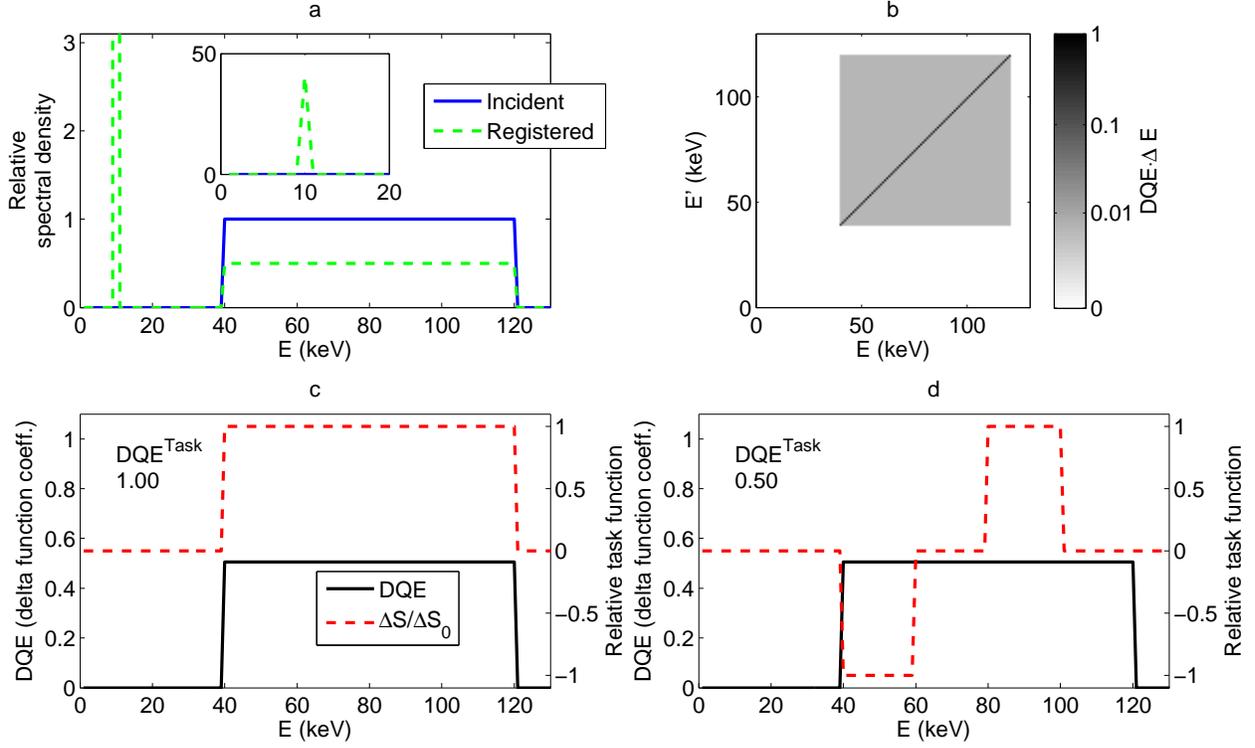}
  \caption{Results for the scatter model. (a) Incident and deposited spectra. Insert: zoomed-out view showing the Compton peak. (b) $\ma{DQE}$ matrix, discretized with $\Delta E=1\; \mathrm{keV}$. Note that the color scale is logarithmic, to visualize the low-intensity background of $6.2\cdot10^{-3}\; \ma{keV}^{-1}$. (c) diagonal of the $\ma{DQE}$ matrix together with the task function and $\mathrm{DQE}^\mathrm{task}$ for a density imaging task. (d) diagonal of the $\ma{DQE}$ matrix together with the task function and $\mathrm{DQE}^\mathrm{task}$ for a spectral imaging task. Since $\ma{DQE}$ is singular, the plotted DQE curves in (c-d) show the coefficient in front of $\delta(E-E')$.}
  \label{fig:scatterToyModel}
\end{figure}

The results for the Compton scatter model are shown in Fig. \ref{fig:scatterToyModel}. Fig. \ref{fig:scatterToyModel}(a) shows the incident and deposited spectra. The $\ma{DQE}$ matrix was discretized by replacing the delta function with a square of width $1\; \mathrm{keV}$. Fig. \ref{fig:scatterToyModel}(c-d) show task functions for (c) density imaging and (d) spectral imaging, together with the diagonal of the DQE matrix and the resulting $\mathrm{DQE}^\mathrm{task}$.

\subsection{CdTe simulation}
\label{sec:results_cdte_simulation}
\begin{figure}
  \centering
  \includegraphics[scale=0.70]{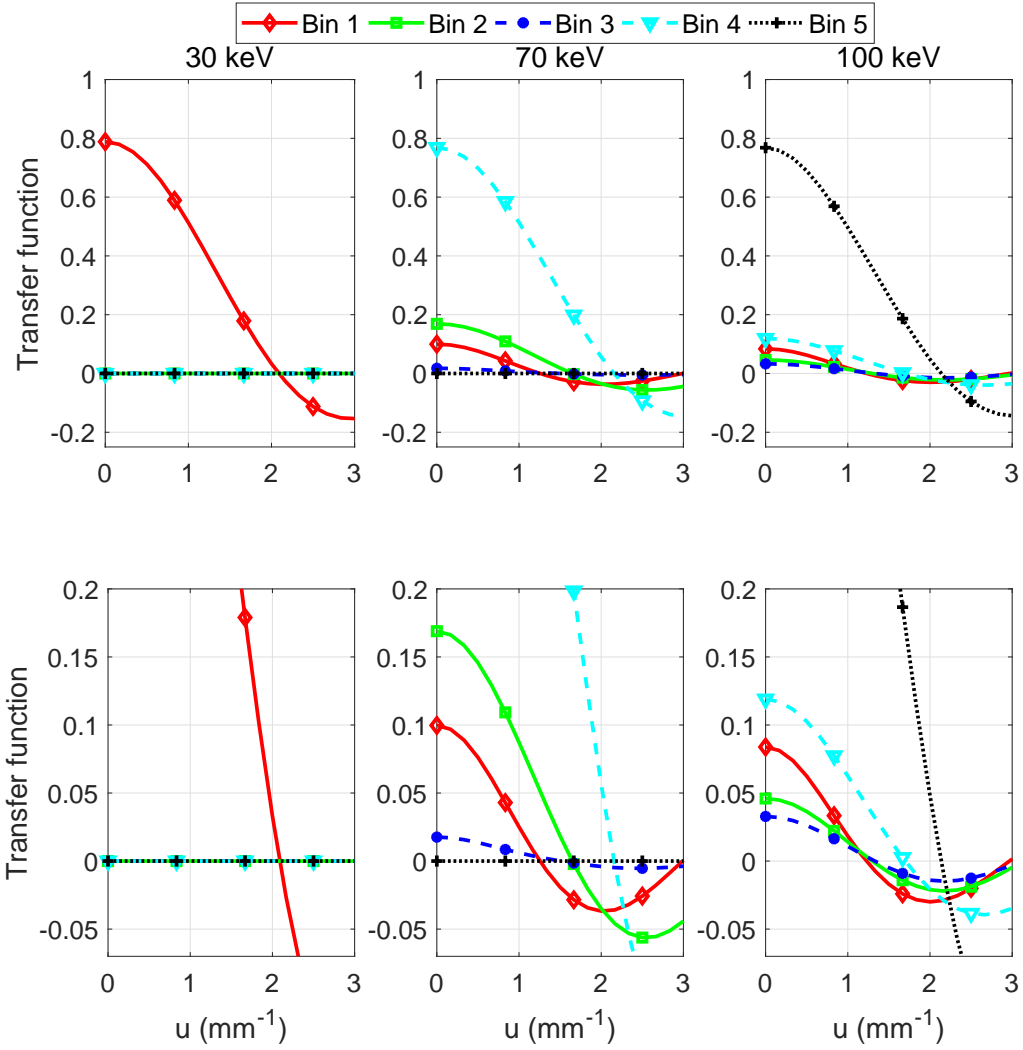}
  \caption{Transfer functions $H_k(\ve{u},E)$ in the CdTe model for three incident monoenergies: 30, 70 and 100 keV. These are plotted along the positive $u$ axis up to three times the Nyquist frequency $1 \; \mathrm{mm}^{-1}$. The lower row contains close-up views of the upper plots.}
  \label{fig:trf}
\end{figure}

\begin{figure}
  \centering
  \includegraphics[scale=0.70]{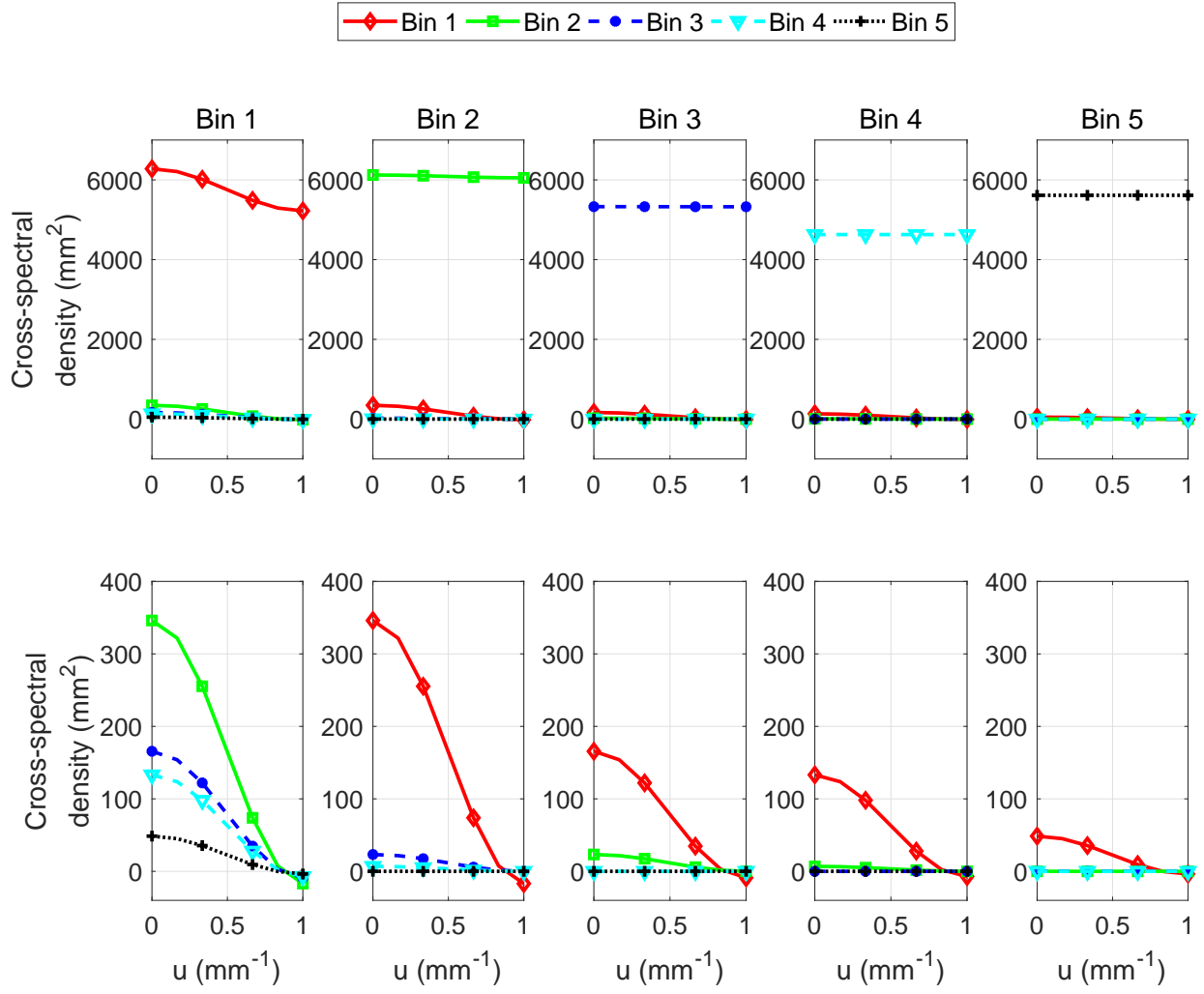}
  \caption{Cross-spectral density $\Delta_x^2\Delta_y^2(W_{d^+})_{kk'}(\ve{u})$ in the CdTe model for broad-spectrum illumination, along the positive $u$ axis up to the Nyquist frequency ($1 \; \mathrm{mm}^{-1}$). Each column of plots represents one $k$ and each curve one $k'$. The lower row contains close-up views of the upper plots. Note that each cross-term is plotted twice since $(W_{d^+})_{kk'}(\ve{u})$ is symmetric in $k$ and $k'$.}
  \label{fig:csd}
\end{figure}

Fig. \ref{fig:psf} shows the simulated point-spread functions for the CdTe model for monochromatic beams of 40, 70 and 100 keV. The corresponding transfer functions $H_k(\ve{u},E)$ are shown in Fig. \ref{fig:trf} for $\ve{u}$ ranging from $u=0 \; \mathrm{mm}^{-1}$ to three times the Nyquist frequency. The matrix elements of the cross-spectral density are shown as a function of frequency in Fig. \ref{fig:csd}, for $u$ from $0 \; \mathrm{mm}^{-1}$ to the Nyquist frequency. Since the autocorrelation function is mirror-symmetric in this model, the cross-spectral density is also symmetric, i.e. extends by mirroring in zero and in the Nyquist frequency.

\begin{figure}
  \centering
  \includegraphics[scale=0.70]{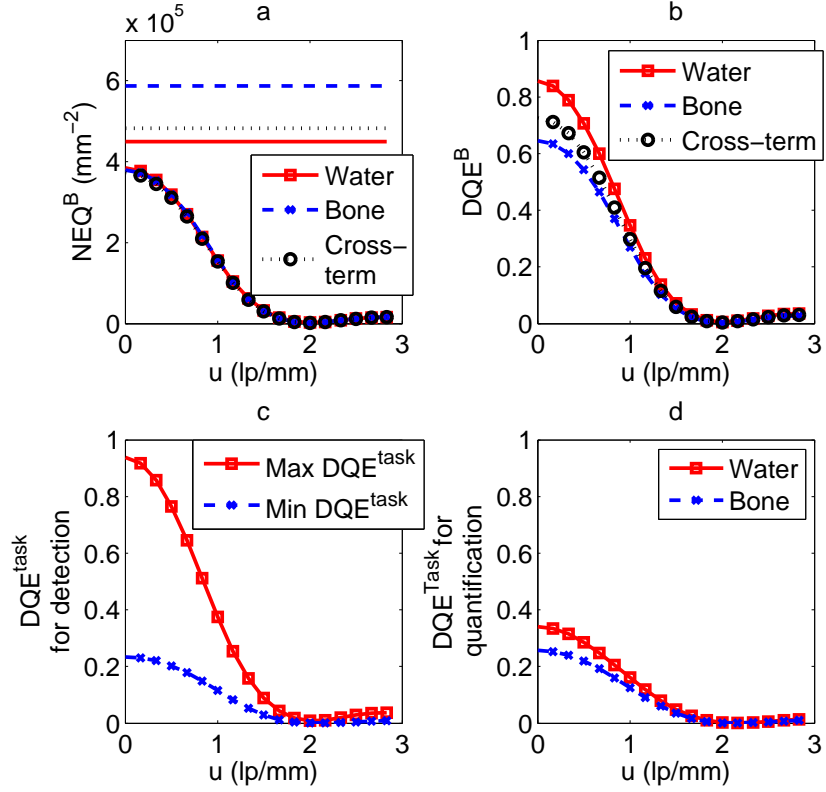}
  \caption{(a) Components of the $\ma{NEQ^{\mathcal{B}}}$ matrix with water and bone as basis functions. The plots with markers apply to the CdTe simulation model whereas the plots without markers are the corresponding values for an ideal detector. (b) Components of the $\ma{DQE^{\mathcal{B}}}$ matrix for the CdTe simulation model with water and bone as basis functions. (c) Maximum and minimum $\mathrm{DQE}^{\mathrm{task}}$ for any detection task, for the CdTe model. (d) $\mathrm{DQE}^{\mathrm{task}}$ for quantifying the amount of water and bone, respectively, in a two-basis decomposition with the CdTe model.} 
  \label{fig:NEQMatBasis}
\end{figure}

The zero-frequency $\ma{NEQ}$ and $\ma{DQE}$ matrices in the basis of monoenergies are displayed in Fig. \ref{fig:NEQEnergyBasis} together with plots of their respective diagonals. Fig. \ref{fig:NEQMatBasis}(a) shows the components of the $\ma{NEQ^{\mathcal{B}}}$ matrix for the CdTe model and for an ideal detector, as a function of spatial frequency for basis materials water and bone. The unattenuated photon fluence of $4\cdot10^6 \; \mathrm{mm}^{-2}$ before the object gives $4.35\cdot10^5 \; \mathrm{mm}^{-2}$ after the object. The corresponding $\ma{DQE^\mathcal{B}}$ is shown for the same basis functions in Fig. \ref{fig:NEQMatBasis}(b), whereas Fig. \ref{fig:NEQMatBasis}(c) shows the largest and smallest $\mathrm{DQE}^\mathrm{task}$ for any detection task. Finally, Fig. \ref{fig:NEQMatBasis}(d) shows $\mathrm{DQE}^\mathrm{task}$ for quantifying water and bone in a two-material decomposition, calculated with Eq. \ref{eq:dqe_task_quantification}.
%discarded: shows the DQE matrix components in the basis of eigenvectors to the CdTe NEQ matrix. Note that these eigenvectors vary with spatial frequency, so the plotted DQE values for different $u$ are expressed in different bases.

\section{Discussion}
\label{sec:discussion}
\ch{In this section, we comment on the results obtained for the fluorescence and scatter toy models and for the CdTe simulation model (Sec. \ref{sec:discussion_fluorescence_model}-\ref{sec:discussion_cdte_simulation}). Then, in Sec. \ref{sec:discussion_applicability_of_the_framework} we discuss the validity and limitations of the proposed theoretical framework.}
\subsection{Fluorescence model}
\label{sec:discussion_fluorescence_model}
\begin{figure}
  \centering
  \includegraphics[scale=0.40]{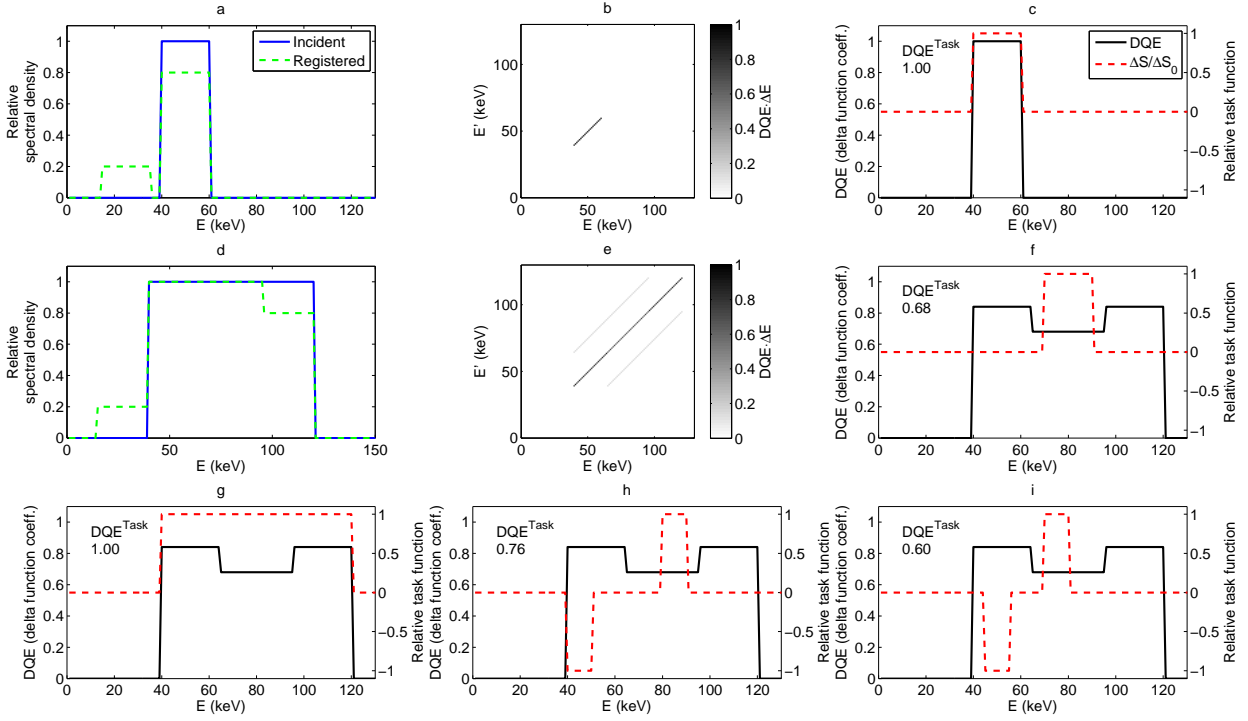}
  \caption{Results for the fluorescence model. (a) Incident and deposited spectra in the nonoverlapping spectrum case, discretized with $\Delta E=1\; \mathrm{keV}$. (b) $\ma{DQE}$ matrix in the nonoverlapping spectrum case. (c) diagonal of the $\ma{DQE}$ matrix together with the task function and $\mathrm{DQE}^\mathrm{task}$ for a density imaging task in the nonoverlapping spectrum case. (d) Incident and deposited spectra in the overlapping spectrum case. (e) $\ma{DQE}$ matrix in the overlapping spectrum case. (f-i) Diagonal of the $\ma{DQE}$ matrix in the overlapping spectrum case, together with task function and $\mathrm{DQE}^\mathrm{task}$ for four different tasks: (f) nonoverlapping density imaging task; (g) overlapping density imaging task; (h) nonoverlapping spectral imaging task; and (i) overlapping spectral imaging task. Since $\ma{DQE}$ is singular, the plotted DQE curves in (c,f-i) show the coefficient in front of $\delta(E-E')$.}
  \label{fig:flourToyModel}
\end{figure}

In the simple fluorescence model, the NEQ matrix (\ref{eq:fluorescence_model}) and the corresponding DQE matrix are independent of spatial frequency, nonzero along the diagonal for the energies contained in the measured spectrum and, if there is spectral overlap, along two off-diagonal lines (Fig. \ref{fig:flourToyModel}(b,e)). To interpret this result, note that according to Eq. \ref{eq:d2_fcn_of_NEQ} the detectability of a feature with differential signal $\Delta S(\ve{u},E)$ at frequency $(\ve{u})$ is obtained as a sum of contributions $\Delta S(\ve{u},E)^* \mathrm{NEQ}(\ve{u},E,E') \Delta S(\ve{u},E)$ from all \textit{pairs} of energies $E,E'$. The first term of (\ref{eq:fluorescence_model}), which contains the diagonal elements of the matrix, shows that a relative signal difference at energy $E$ contributes to $d'^2$ with two terms. The first term is the contribution from the non-fluorescence (photopeak) interactions at energy $E$ and the second term is the contribution from the fluorescence interactions with original energy $E$ and registered energy $E-E_F$. The two off-diagonal contributions in (\ref{eq:fluorescence_model}) are nonzero only for $E'=E+E_F$ and $E'=E-E_F$, respectively. These terms contain the contribution from overlap between the photopeak and the K-escape peak. The denominator for each of these contributions is the quantum noise variance, equal to the number of registered events at the deposited energy.

When the incident spectrum is narrow enough that the photopeak and K-escape spectra do not overlap (Fig. \ref{fig:flourToyModel}(a-c)), there are no off-diagonal terms and the $\ma{DQE}$ matrix is a delta function along the diagonal with a coefficient of 1. Even though the K-escape photons are registered with the wrong energy, there is no risk that these will be confused with photopeak photons, and since there is no ambiguity about the true energy, the K-escape photons do not cause any performance degradation at all and the $\mathrm{DQE}^{\mathrm{task}}$ for any imaging task is 1.

When the spectrum is broad enough to give overlap between the photopeak and K-escape events, on the other hand, off-diagonal elements appear in the $\ma{DQE}$ matrix and the values of the diagonal elements decrease (Fig. \ref{fig:flourToyModel}(d-i)). The diagonal elements of the DQE matrix indicate how much $\Delta S(\ve{u},E)$ at each energy $E$ alone contribute to the total $d'^2$, regardless of what other energies are present in $\Delta S(\ve{u},E)$. In this case, the diagonal DQE elements are highest near the maximum energy, where the recorded spectrum does not include any K-escape component, and near the minimum energy, where the K-escape events from these photons are registered below the minimum incident energy and do not have to compete with the noise from photopeak events.

The off-diagonal terms, on the other hand, specify the contribution to $d'^2$ caused by the presence of a signal at both energy $E$ and $E'$. This contribution can be either positive or negative depending on the task. Since the off-diagonal terms are positive in this model, the detectability increases if $\Delta S$ has the same sign at both energies and decreases if $\Delta S$ has opposite signs at the two energies. A case when $\Delta S$ has the same sign at $E$ and $E'$ can correspond e.g. to a density imaging task, where the total number of registered photons is more important than the energy distribution. In this case the simultaneous presence of photons at both energies in $\Delta S$ is actually beneficial, since the $d'^2$ contribution is proportional to the square \ch{of} the differential signal, and increasing spectral overlap in the signal (while keeping the background noise fixed) helps concentrate more signal at a single measurement energy. \ch{Note, however, that $d'^2$ can never exceed the value that would have been obtained if all photons were registered correctly on the first place.} On the other hand, a case where $\Delta S$ has opposite signs at the two energies is a material discrimination task, e.g. K-edge imaging. Here, the simultaneous presence of energies $E$ and $E'$ decreases the detectability since spectral overlap degrades the ability of the detector to measure true spectral differences. 
%discarded: and inversely proportional to the noise 

In the model studied here, the off-diagonal terms are nonzero only for $E-E' = E_F$. The impact of the fluorescence on detectability is therefore dependent on whether the task function $\Delta S (\ve{u},E)$ involves energies with spectral overlap or not. This is demonstrated in Fig. \ref{fig:flourToyModel}(f-i). When the task is to detect a signal difference in a narrow energy band, the $\mathrm{DQE}^{\mathrm{task}}$ is simply given by the diagonal value of the $\ma{DQE}$ matrix for that energy range (Fig. \ref{fig:flourToyModel}(f)). This is less than unity because the registered change in photon flux must be detected in the presence of a larger noise contribution from co-registered K-escape and photopeak events. On the other hand, when the task is to detect a signal difference across a broad range of energies, the off-diagonal elements cause constructive interference and $\mathrm{DQE}^{\mathrm{task}}=1$, as for an ideal detector (Fig. \ref{fig:flourToyModel}(g)). For a density imaging task, the off-diagonal elements thus improve $\mathrm{DQE}^{\mathrm{task}}$.
%Discarded: In the model studied here, only the diagonal DQE elements contribute to $d'^2$ if $E-E' \ne E_F$, but if $E-E' = E_F$ a contribution appears that can be either positive or negative depending on whether $\Delta S$ has the same or different sign at the two energies. 

For a spectral imaging task, i.e. when trying to detect an attenuation difference between two energy bands, the off-diagonal elements have a different effect (Fig. \ref{fig:flourToyModel}(h-i)). For a spectral task where the spectral bands do not have any overlap due to fluorescence, $\mathrm{DQE}^{\mathrm{task}}$ is given by the average over the two energy bands of the delta function coefficient along the diagonal of $\ma{DQE}$, i.e. the off-diagonal elements do not affect the result. If the bands are positioned so that K-escape events from one of them overlaps with photopeak events from the other, the off-diagonal elements give negative interference and decrease $\mathrm{DQE}^{\mathrm{task}}$. In this case, the off-diagonal elements are detrimental, since cross-talk from one energy band to the other makes it more difficult to detect a spectral difference.

%Discarded: If the NEQ matrix is formulated in another basis, a similar interpretation of its elements can be made, in terms of the contributions from the presence of different bases in the signal difference.
\subsection{Scatter model}
\label{sec:discussion_scatter_model}
In the Compton scatter model (Fig. \ref{fig:scatterToyModel}) half the incident spectrum is registered at its correct energy, while half is registered as a peak at 10 keV (Fig. \ref{fig:scatterToyModel}(a)) This is reflected in the form of the DQE matrix (Fig. \ref{fig:scatterToyModel}(b)), which is the sum  of two contributions. The photoelectric interactions give a diagonal with delta functions whose coefficient is 0.5, i.e. half the DQE of an ideal detector. The Compton interactions provide information that a photon was detected but do not contain any energy information. Their contribution is evenly distributed as a weak background over the entire square formed by the spectral supports of $\overline{q}(E)$ and $q(E')$. The total area of the diagonal line and the background is the same since each of them are generated from $50\%$ of the photons.

For a density imaging task where $\Delta S$ is constant as a function of energy, $d'^2$ is given by the integral over nonzero part of the $\ma{DQE}$ matrix, so that the two contributions add and give $\mathrm{DQE^{task}}=1$. Since only the total number of registered photons matters in this case, the fact that some of them are registered at the wrong energy does not degrade the performance. For a spectral imaging task on the other hand, the detectability is  obtained as an integral  of the NEQ over four squares in the $(E,E')$ diagram, two near the diagonal with positive sign and two away from the diagonal with negative sign. The uniform background term in the NEQ matrix therefore gives a net contribution of 0, so that $\mathrm{DQE^{task}}=0.5$, reflecting only the photoelectric contribution. The Compton events, which are all registered with the same energy, thus contain as much information as photoelectric events for density imaging tasks but no information at all for a pure spectral task. A real-world imaging task will typically be an intermediate between these two types, meaning that the Compton events will contribute some information but less than the photoelectric events.  Also note that this model is simplified in the sense that, in a real detector, the deposited energy from a Compton interaction is random with a distribution function that depends on the incident energy and the scattering angle, meaning that the Compton interactions do contain some energy information.
%discarded, mentioned elsewhere: This model is simplified in the sense that we have ignored that photons will typically be reabsorbed at another site in the sensor, degrading spatial resolution and generating noise correlations. 

\subsection{CdTe simulation}
\label{sec:discussion_cdte_simulation}
\begin{figure}
  \centering
  \includegraphics[scale=0.70]{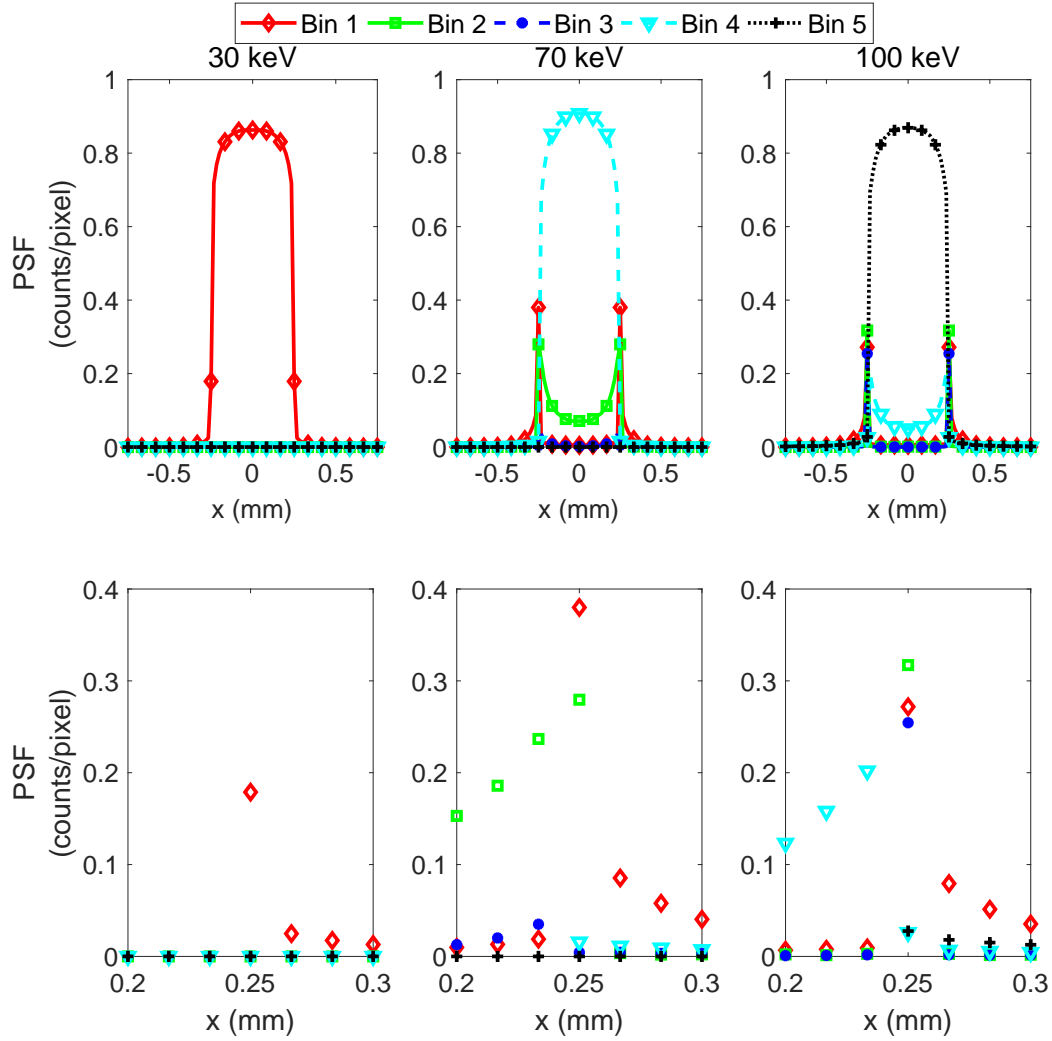}
  \caption{Slices along the $x$ axis of the simulated energy-bin point-spread functions in the CdTe model measured as \ch{counts} per pixel, $\Delta_x\Delta_yh_k(\ve{\Delta r},E)$, for three incident monoenergies: 30, 70 and 100 keV. The lower row contains close-ups of the region closest to the pixel border. (In the lower row, the data points are not connected by lines since the psf is not expected to be smooth.)}
  \label{fig:psf}
\end{figure}

As shown in Fig. \ref{fig:psf}, the point-spread function $h_k(\ve{\Delta r},E)$ of the CdTe simulation model depends on the incident energy. When a monoenergetic beam impinges on the interior of a pixel, the majority of counts are registered in the energy bin that includes the true energy: bin 1, 4 and 5 for 30, 70 and 100 keV, respectively. For a 30 keV beam, nothing is registered in the other bins, since pileup is not included in the model and the energies of different photons thus cannot be added together. For higher incident energies, some photons are misregistered in lower energy bins due to charge sharing and fluorescence. For example, a 100 keV beam incident precisely on the border between two pixels will be registered in energy bin 2 (41-54 $\mathrm{keV}$) since only about half the photon energy, 50 keV, is deposited in the studied pixel. If the same beam hits inside the pixel but close to the border, a fraction (up to $20\%$) of the photons are registered in bin 4 (65-77 $\mathrm{keV}$) since they only lose a minor part of their energy to charge sharing or fluorescence. Finally, just outside the pixel border, the majority of the events are registered in energy bin 1 (25-40 $\mathrm{keV}$) since most of the energy of these photons is deposited in the neighboring pixel.

The shape of the point-spread function is reflected in the transfer function $H_k(\ve{u},E)$ (Fig. \ref{fig:trf}). In each case, the transfer function for the energy bin encompassing the incident energy falls off from a zero-frequency value close to 0.8, signifying that about 80\% of the total number of photons are registered in that energy bin, and crosses 0 near twice the Nyquist frequency, reflecting the nearly rectangular point-spread function. The other energy bins detect smaller fractions of the photons and fall off more rapidly, indicating that the blurring from the inter-pixel cross-talk suppresses high spatial frequencies.

The low-pass character of the cross-talk can also be seen from the cross-spectral density plotted in Fig. \ref{fig:csd}. As seen in these figures, the diagonal elements of $W_{d^+}$ are nearly constant functions of spatial frequency, which indicates that the noise in these is uncorrelated between pixels. The exception is the lowest energy bin, which has a low-frequency component in its noise stemming from the fact that some photons are counted in this energy bin in more than one pixel. The cross-elements of $W_{d^+}$ exhibit strong low-frequency correlations since photons are typically registered in one energy bin in the pixel of incidence and in another energy bin in a neighboring pixel. The strongest such correlation is found between energy bins 1 and 2. This shows that the correlation structure in the measured image requires a joint description in terms of both spatial and energy correlations.

\begin{figure}
  \centering
  \includegraphics[scale=0.70]{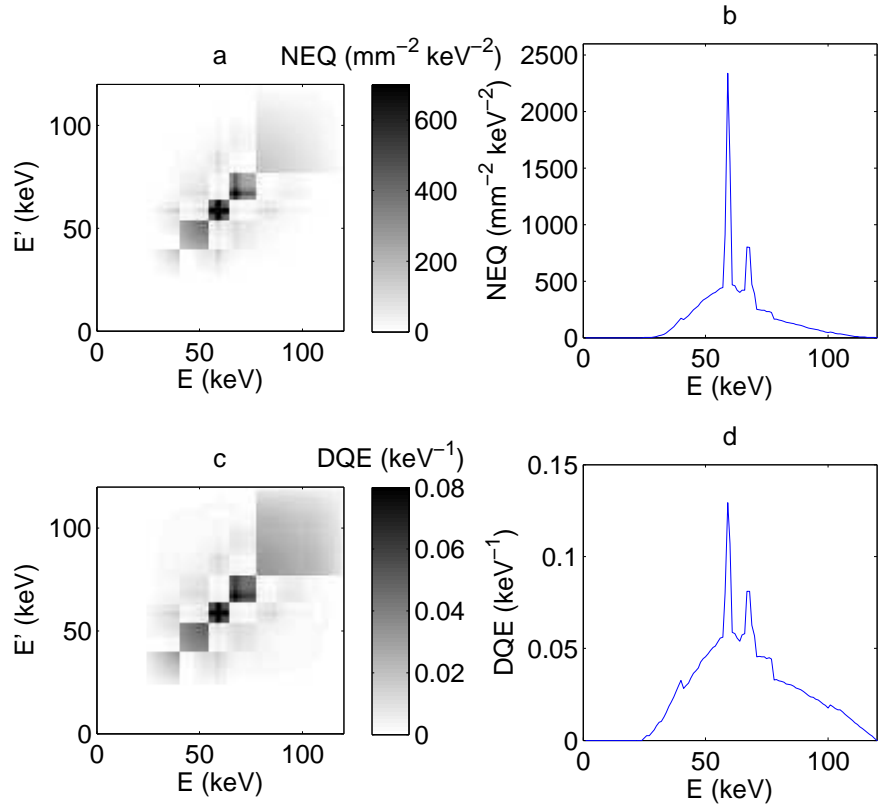}
  \caption{(a) Zero-frequency $\ma{NEQ}$ matrix for the CdTe model. (b) Diagonal elements of the $\ma{NEQ}$ matrix in (a). (c) Zero-frequency $\ma{NEQ}$ matrix for the CdTe model. (d) Diagonal elements of the $\ma{DQE}$ matrix in (c).}
  \label{fig:NEQEnergyBasis}
\end{figure}

The zero-frequency $\ma{NEQ}$ and $\ma{DQE}$ matrices in the basis of monoenergies (Fig. \ref{fig:NEQEnergyBasis}) exhibit a block-like structure with the block borders corresponding to the energy thresholds. These plots show that the largest cross-terms, and thereby the strongest coupling between different energies, is found within each energy bin, i.e. in the diagonal blocks. Note that the relative intensity of the blocks resembles the three diagonal bands caused by the loss of a fixed amount of energy in the simplified fluorescence model (Fig. \ref{fig:flourToyModel}(e)). Figs \ref{fig:NEQEnergyBasis}(b,d) also show that both the NEQ and DQE are largest for the energies corresponding to the characteristic X-ray peaks in the spectrum. All energies falling within a certain energy bin are detected against the same amount of background noise, and since the detectability is a quadratic function of the signal difference, a certain relative signal difference $\Delta S$ is gives a larger detectability when the spectral density is higher. It is therefore easier to detect an attenuation difference right at a characteristic X-ray energy of the source than at a neighboring energy. This also explains why the matrix-valued DQE is dependent on the X-ray spectrum shape.

The $\ma{NEQ}^{\mathcal{B}}$ and $\ma{DQE}^{\mathcal{B}}$ are plotted as functions of spatial frequency in Fig. \ref{fig:NEQMatBasis}(a-b). Note that the NEQ values for the ideal detector are slightly higher than the transmitted photon fluence of $4.35\cdot10^5 \; \mathrm{mm}^{-2}$, in particular for bone, reflecting the higher detectability of an ideal energy-resolving detector compared to an ideal photon-counting detector with no energy discrimination. The task-specific $\ma{DQE}$ values for detecting water and bone are obtained directly from Fig. \ref{fig:NEQMatBasis}(b) as the diagonal elements of the $\ma{DQE}^{\mathcal{B}}$ matrix. The plots show that the performance for both detection of bone and soft tissue suffer relatively little from the degraded energy resolution caused by charge sharing and fluorescence, since the zero-frequency $\mathrm{DQE^{task}}$ is 0.86 for water and 0.65 for bone. The sinc-like decrease towards higher frequencies is caused by the pixel aperture. The cross-term is positive, meaning that the presence of both soft tissue and bone in a detection task will cause constructive interference, i.e. increase detectability. 

The maximum $\mathrm{DQE^{task}}$ for any imaging task is 0.94, while the minimum is 0.23 (Fig. \ref{fig:NEQMatBasis}(c)). Both water and bone detection are therefore tasks for which the performance of the studied detector comes relatively close to the maximum. For material quantification, on the other hand, the $\mathrm{DQE^{task}}$ is severely degraded by the imperfect energy resolution of the detector (0.34 for water and 0.26 for bone) and comes close to its lowest possible value for any detection task. (Fig. \ref{fig:NEQMatBasis}(d)). Thus, the mis-registration of photon energies has a limited effect on detecting a small change in the amount of water or bone, since this also affects the total number of photons, but a large effect on the ability to determine what material caused a loss of flux, which requires good energy resolution.
%NOTE I've thought about relating this to synthetic monoenergetic imaging but haven't found an obvious relationship. How does this work with frequency dependence?

In the model studied here, we have neglected electronic noise for the purpose of making the effects of the spectrum shape and nonideal energy response stand out clearly. In a real detector, electronic noise will blur the detected spectrum, so that the transition of the point-spread function at the pixel border becomes smoother compared to Fig. \ref{fig:psf} and the characteristic X-rays are less prominent in the $\ma{NEQ}$ matrix compared to Fig. \ref{fig:NEQEnergyBasis}.

\subsection{\ch{Applicability of the framework}}
\label{sec:discussion_applicability_of_the_framework}
% Discarded:
% Finally, we will address the question 
\ch{Since a completely realistic detector model would be mathematically intractable, the proposed framework builds on a number of idealized assumptions. For the detectability formula in basis format (\ref{eq:d2_fcn_of_NEQB}) to be valid, the feature to be detected must be weakly attenuating, i.e. either small or low-contrast. The detector response is assumed to be linear and shift invariant, and the noise is assumed to be stationary. These assumptions are not true in general in CT, since the photon flux can vary greatly between different projection lines. This makes the noise non-stationary and can give pileup in some projections, meaning that the response will be nonlinear and position-dependent. However, our model is valid in regions of the image where the photon flux is slowly varying, so that the detector response is approximately linear. The primary usefulness of this framework therefore lies in its ability to aid understanding and optimization of detector performance for model tasks where a feature is to be detected against a slowly varying background. On the other hand, imaging of phantoms, in either simulation or experiment, will be needed in order to measure the performance for more complex tasks. Also note that the assumption that aliasing is negligible means that one must use caution when using the framework to predict detection performance for high-frequency tasks.}

\ch{There are also effects that are not included in the work presented here but could be included in the future. One example is the impact of scatter from the object that is not included here but can be modeled as additive noise. Also, our present model does not include pile-up and electronic readout noise, but these effects can be taken into account by future extensions of the framework.}

\ch{The ideal-linear-observer performance derived here may not always mimic the performance of a human observer. However, if the measured count numbers are large enough to be well described by Gaussian statistics, the optimal-linear-observer performance equals the performance of the ideal observer\cite{barrett_myers_foundations} and} \chtwo{can therefore be expected to give} \ch{an approximate upper limit to the detection performance achievable with advanced image processing.} %Moved from conlusion but unchanged

%Discarded: Another important question is to what extent the NEQ and DQE matrices give an accurate description of the imaging performance when reconstruction or image processing algorithm is applied to the raw data. If this post-processing algorithm is a linear, shift-invariant transformation, the Fourier transformed signal will be transformed as 
\chtwo{To further investigate the meaning of the NEQ and DQE matrices when a reconstruction or image processing algorithm is applied to the raw data, we start by studying the case where the post-processing algorithm is a linear, shift-invariant transformation. In this case,} \ch{the Fourier transformed signal will be $\overline{D}^{\mathrm{postproc}}_k(\ve{u})=\sum_l \chtwo{\chtwo{U}}_{kl}(\ve{u})\overline{D}_l(\ve{u})$ where $\chtwo{U}_{kl}(\ve{u})$ are the components of a frequency-dependent transformation matrix. The combination of the detector and the algorithm can be seen as a composite detection system with a transfer function $H^{\mathrm{sys}}_k(\ve{u},E)=\sum_l \chtwo{U}_{kl}(\ve{u})H_k(\ve{u},E)$ and cross-spectral density $\ma{W}^{\mathrm{sys}}_{d^+}(\ve{u})$ fulfilling $\left[W^{\mathrm{sys}}_{d^+}\right]_{kk'}(\ve{-u})=\sum_{ll'} \chtwo{U}_{kl}(\ve{u})\left[W_{d^+}\right]_{ll'}(\ve{-u})\chtwo{U}^*_{k'l'}(\ve{u})$.  (The latter identity follows from Eq. \ref{eq:appendix_fourier_cov_components} and the corresponding formula for $\left[W^{\mathrm{sys}}_{d^+}\right]_{kk'}(\ve{-u})$ with $\ve{d}^s$ replaced by $\fourop_{\mathrm{DS}}^{-1}\chtwo{U}(\ve{u})\fourop_{\mathrm{DS}}\ve{d}^s$.) Now, Eqs. \ref{eq:neq_expressed_in_H_and_W}, \ref{eq:dqe_definition} and \ref{eq:CRLB_and_NEQ} show that this leaves the frequency-dependent NEQ and DQE matrices as well as the frequency-dependent CRLB unchanged. These performance metrics are thus invariant under all linear shift-invariant processing algorithms and are therefore well suited for analyzing imaging performance. Also note that the transfer function $H^\mathrm{sys}_k(\ve{u},E)$ and cross-spectral density $\ma{W}_{d+}^\mathrm{sys}(\ve{u})$ of the composite system are useful quantities in their own right for studying the effect of linear postprocessing algorithms, e.g. spectral distortion correction or deblurring algorithms.}

\ch{In contrast to linear algorithms, nonlinear image processing algorithms, such as many iterative reconstruction\cite{thibault_statistical_iterative} and denoising\cite{brown_anticorrelated} algorithms, cannot be readily described by the present framework, although they may in some cases be linearized and thereby analyzed in an approximative sense. The NEQ and DQE metrics should thus be regarded as measures of how much information is available in the raw output data from the detector, i.e. they provide information about the amount of information available as input to the algorithm. Under the reasonable assumption that providing more input information to most such algorithms will also lead to better output images, the proposed metrics can be expected to be useful predictors of image quality also when nonlinear processing is applied. However, further investigations will be necessary in order to establish the validity of this assumption.}

\section{Conclusion}
\label{sec:conclusion}
In this work, we have shown how the framework of linear-systems theory can be extended to describe energy-resolving detectors. We have demonstrated a natural way of generalizing the NEQ and DQE metrics to matrix-valued quantities containing information about both the spatial resolution, detection efficiency and energy resolution of the detector. Furthermore, we have demonstrated how basis material decomposition can be used to express these matrices in a compact form and that they are closely related to \ch{a generalized version of} the Cram\'{e}r-Rao lower bound which describes the detector performance for material decomposition. We have thus merged two approaches for detector performance assessment, the linear-systems framework for describing detection task performance and the CRLB approach for describing material decomposition performance.

Although photon-counting detectors have been studied in the examples presented here, the proposed framework can also be extended to other spectral imaging systems, such as dual-layer detectors. \ch{In future work, we will also use the present framework to study more realistic detector systems, where pileup and electronic noise are taken into account.} Another topic for future work is to develop methods for measuring the matrix-valued NEQ and DQE experimentally, for detector modules and for complete imaging systems. This new framework for detector performance characterization will help the development of photon-counting X-ray imaging systems by facilitating comparisons between different detector designs and elucidating the trade-off between different parameters, such as spatial resolution, energy resolution and dose efficiency.

% use section* for acknowledgement
\section*{Acknowledgments}
This study was supported by NIH Grant U01 EB01714003. The authors would like to thank Moa Yveborg and Jesse Tanguay for helpful discussions. 
\section*{Disclosure of Conflicts of Interest}
Mats Persson is stockholder of and consultant for Prismatic Sensors AB. Norbert J. Pelc is consultant for Prismatic Sensors AB.

\appendix
\section{Fourier-domain covariance matrix}
\label{sec:appendix_matrix_transformation}
In this section we derive the Fourier-domain expression for the optimal-linear-observer detectability. The matrix elements of the discrete-space Fourier operator and its inverse is $\left(\fourop_{\mathrm{DS}}\right)_{\ve{n}}(\ve{u})=\mathrm{e}^{-2\pi \imu\ve{u}\cdot\ve{r_n}}$ and $\left(\fourop_{\mathrm{DS}}^{-1}\right)_{\ve{n}}(\ve{u})=\Delta_x\Delta_y\mathrm{e}^{2\pi \imu\ve{u}\cdot\ve{r_n}}$. Eq. \ref{eq:d2_spatial_domain} then gives
\begin{equation}
  \begin{aligned}
    \label{eq:appendix_d2_start}
    d'^2 =\left(\fourop_{\mathrm{DS}}\ve{\Delta \overline{d}}^s\right)^\dagger \left(\fourop_{\mathrm{DS}}^\dagger\right)^{-1}\mathrm{Cov}(\ve{d}^s)^{-1}\fourop_{\mathrm{DS}}^{-1} \fourop_{\mathrm{DS}}\ve{\Delta \overline{d}}^s\\
    = \left(\ve{\Delta \overline{D}}^s\right)^\dagger \left(\fourop_{\mathrm{DS}}\mathrm{Cov}(\ve{d}^s)\fourop_{\mathrm{DS}}^{\dagger}\right)^{-1} \ve{\Delta \overline{D}}^s,\\
  \end{aligned}
\end{equation}
where $\dagger$ denotes conjugate-transpose. Since $\mathrm{\ve{d}^s}$ is wide-sense stationary,

%Old formulas
%= \left(\Delta \overline{D}^s\right)^\dagger \Delta_x\Delta_y\fourop_{\mathrm{DS}}\mathrm{Cov}(d^s)^{-1}\fourop_{\mathrm{DS}}^{-1} \Delta \overline{D}^s\\
%= \left(\Delta \overline{D}^s\right)^\dagger \Delta_x\Delta_y\left(\fourop_{\mathrm{DS}}\mathrm{Cov}(d^s)\fourop_{\mathrm{DS}}^{-1}\right)^{-1} \Delta \overline{D}^s\\
\begin{equation}
  \label{eq:appendix_fourier_cov_components}
  \begin{aligned}
    \left[\fourop_{\mathrm{DS}}\mathrm{Cov}(\ve{d}^s)\fourop_{\mathrm{DS}}^{\dagger}\right]_{k,k'}\!\!(\ve{u},\ve{u}')
    =\sum_{\ve{n}=-\infty}^{\infty}\sum_{\ve{n}'=-\infty}^{\infty} \mathrm{Cov}(d^s_{\ve{n},k},d^s_{\ve{n}',k'})\mathrm{e}^{2\pi \imu(-\ve{u}\cdot\ve{r_n}+\ve{u'}\cdot\ve{r}_{\ve{n}'})}\\
    =\sum_{\ve{n}=-\infty}^{\infty}\sum_{\Delta \ve{n}=-\infty}^{\infty} K^s_{\ve{\Delta n},k,k'}\mathrm{e}^{2\pi \imu((\ve{u}'-\ve{u})\cdot \ve{r_n}+\ve{u'}\cdot \ve{\Delta r_n}}
    =\sum_{\ve{n}=-\infty}^{\infty}\mathrm{e}^{2\pi \imu(\ve{u}'-\ve{u})\cdot\ve{r_n}} \sum_{\ve{\Delta n}=-\infty}^{\infty} K^s_{\ve{\Delta n},k,k'}\mathrm{e}^{2\pi \imu\ve{u'}\cdot \ve{\Delta r_n}}\\
    =\sum_{\ve{n}=-\infty}^{\infty}\mathrm{e}^{2\pi \imu(\ve{u}'-\ve{u})\cdot\ve{r_n}} W^s_{k,k'}(-\ve{u}')
    =\frac{1}{\Delta_x\Delta_y}\delta(\ve{u}-\ve{u}') W^s_{k,k'}(-\ve{u}').
  \end{aligned}
\end{equation}
\ch{within the Nyquist region $\mathrm{Nyq}=\left\{\ve{u}: \left|u\right|<\frac{1}{2\Delta_x}, \left| v\right|<\frac{1}{2\Delta_y}\right\}$.} Here, $\ve{\Delta n}=\ve{n}'-\ve{n}$ and $\ve{\Delta r_n}=\ve{r}_{\ve{n}'}-\ve{r_n}$. Substituting into (\ref{eq:appendix_d2_start}) and exploiting the conjugate-symmetry of $W^s_{k,k'}(\ve{u})$ gives
\begin{equation}
  \begin{aligned}
    \label{eq:appendix_d2_expanded}
    d'^2 = \left(\ve{\Delta \overline{D}}^s\right)^\dagger \left(\fourop_{\mathrm{DS}}\mathrm{Cov}(\ve{d}^s)\fourop_{\mathrm{DS}}^{\dagger}\right)^{-1} \ve{\Delta \overline{D}}^s\\
    = \Delta_x\Delta_y\int_{\mathrm{Nyq}} \sum_k \sum_{k'} \Delta \overline{D}^s_k(\ve{u})^* \left[\left({W^s}\right)^{-1}\right]^*_{k,k'}(\ve{u}) \Delta \overline{D}^s_{k'}(\ve{u})\mathrm{d}\ve{u}.
  \end{aligned}
\end{equation}
\ch{where ${}^*$ denotes complex conjugate.}

\ch{We will now express the noise correlations in terms of $\ma{W}_{d^+}(\ve{u})$, the cross-spectral density of the sampled pulse train signal $d_k^+(\ve{r})=\sum_{\ve{n}=-\infty}^\infty d^s_{\ve{n},k}\delta(\ve{r}-\ve{r_n})$. By observing that $K^s_{\ve{\Delta n},k,k'}=(K_d)_{k,k'}(\ve{\Delta r_n})$, where $K_d$ is the autocovariance of the presampling signal $d$, and using  $(K_{d^+})_{k,k'}(\ve{\Delta r})=\frac{1}{\Delta_x\Delta_y}(K_d)_{k,k'}(\ve{\Delta r})\sum_{\ve{n}}\delta(\ve{\Delta r}-\ve{\Delta r_n})$ (Ref. \citenum{cunningham_linear_systems_chapter}, eq. 2.108), we obtain $W^s_{k,k'}(\ve{u})=\Delta_x\Delta_y(W_{d^+})_{k,k'}(\ve{u})$. Assuming that aliasing is negligible, $\Delta \overline{D}^s_k(\ve{u})\approx\frac{1}{\Delta_x\Delta_y}\Delta \overline{D}_k(\ve{u})$, and using Eq. \ref{eq:signal_difference_as_convolution} then allows us to express (\ref{eq:appendix_d2_expanded}) as}

\begin{equation}
  \begin{aligned}
   \label{eq:d2_q_inserted_appendix}
    \ch{d'^2=  \int_{\mathrm{Nyq}} \sum_k \sum_{k'} \int_0^{E_\mathrm{max}} \int_0^{E_\mathrm{max}} H_k(\ve{u},E)^*\Delta\overline{Q}(\ve{u},E)^*}\\
    \ch{\cdot\left[W_{d^+}^{-1}\right]^*_{k,k'}\!\!(\ve{u}) H_k(\ve{u},E') \Delta\overline{Q}(\ve{u},E')\mathrm{d}E \mathrm{d}E'\mathrm{d}\ve{u}.}
  \end{aligned}
\end{equation}
\ch{which is (\ref{eq:d2_q_inserted}).}

\section{Derivation of the CRLB}
\label{sec:appendix_fisher_matrix}
In this section we derive the form of the CRLB for $\tilde{\ve{A}}(\ve{u})$ in the Fourier domain. \ch{To obtain the CRLB for the material decomposition $\tilde{\ve{A}}$ we introduce a change of variables by defining $\underline{\tilde{\ve{A}}}=\left(\tilde{\ve{A}},\tilde{\ve{A}}^*\right)\trsp=\ma{T}\bm{\uptheta}$. Here $\ma{T}$ is the block matrix\cite{schreier_scharf_2010} $\begin{pmatrix} \ma{I} & i\ma{I}\\ \ma{I} & -i\ma{I} \end{pmatrix}$, where $\ma{I}$ is the identity matrix on the vector space on which $\mathrm{Re}\;\tilde{\ve{A}}, \mathrm{Im}\;\tilde{\ve{A}}$ are defined.} Using Eq. \ref{eq:fisher_first_term} and the identity $\ma{T}^{-1}=\frac{1}{2}\ma{T}^\dagger$ we get the CRLB for $\underline{\tilde{\ve{A}}}$ as
%Discarded
% \begin{equation}
%   \ma{T}=
%   \begin{pmatrix}
%     \ma{I} & i\ma{I}\\
%     \ma{I} & -i\ma{I}
%   \end{pmatrix}
% \end{equation}
\begin{align}
  \mathrm{Cov}\left(\underline{\hat{\tilde{\ve{A}}}}\right)^{-1}=\left(\ma{T}\mathrm{Cov}\left(\bm{\uptheta}\right)\ma{T}^\dagger\right)^{-1}=\frac{1}{4}\ma{T}\left[\mathrm{Cov}\left(\bm{\uptheta}\right)^{-1}\right]\ma{T}^{\dagger}\\ \nonumber
  \le \frac{1}{4}\ma{T}\left(\frac{\partial \ve{\overline{d}}^s}{\partial \bm{\uptheta}}\right)\trsp \mathrm{Cov}(\ve{d}^s)^{-1}\frac{\partial \ve{\overline{d}}^s}{\partial \bm{\uptheta}}\ma{T}^{\dagger}
  =\left(\frac{\partial\ve{\overline{d}}^s}{\partial \tilde{\ve{A}}}, \frac{\partial\ve{\overline{d}}^s}{\partial \tilde{\ve{A}}^*}\right)^\dagger \mathrm{Cov}(\ve{d}^s)^{-1}\left(\frac{\partial\ve{\overline{d}}^s}{\partial \tilde{\ve{A}}} ,\frac{\partial\ve{\overline{d}}^s}{\partial \tilde{\ve{A}}^*}\right)=\underline{\ma{F}}_{\underline{\tilde{\ve{A}}}},
\end{align}
where $\underline{\ma{F}}_{\tilde{\ve{A}}}$ is the Fisher matrix for $\underline{\tilde{\ve{A}}}$. Here the derivatives with respect to $\ve{A}$ and $\ve{A}^*$ are defined as Wirtinger derivatives, which allows differentiating both real and complex functions with respect to a complex variable (See Ref. \citenum{schreier_scharf_2010}, ch. A2). This gives $\underline{\ma{F}}_{\underline{\tilde{\ve{A}}}}=\begin{pmatrix} \ma{F}_{\tilde{\ve{A}}} & \tilde{\ma{F}}_{\tilde{\ve{A}}}\\ \tilde{\ma{F}}_{\tilde{\ve{A}}}^* & \ma{F}_{\tilde{\ve{A}}}^* \end{pmatrix}$ with $\ma{F}_{\tilde{\ve{A}}}=\left(\frac{\partial\ve{\overline{d}}^s}{\partial \tilde{\ve{A}}}\right)^\dagger \mathrm{Cov}(\ve{d}^s)^{-1}\left(\frac{\partial\ve{\overline{d}}^s}{\partial \tilde{\ve{A}}}\right)$ and $\tilde{\ma{F}}_{\tilde{\ve{A}}}=\left(\frac{\partial\ve{\overline{d}}^s}{\partial \tilde{\ve{A}}}\right)^\dagger \mathrm{Cov}(\ve{d}^s)^{-1}\left(\frac{\partial\ve{\overline{d}}^s}{\partial \tilde{\ve{A}}}\right)^*$

Since Eq. \ref{eq:fourier_transform_q_and_a_relation} gives $\frac{\partial \overline{Q}(\ve{u}',E)}{\partial \tilde{A}_l(\ve{u})}=\frac{\partial\overline{q}(E)}{\partial A_l}\delta(\ve{u}'-\ve{u})$ in the linear approximation, we can calculate the Jacobian of $\ve{\overline{d}}^s$ using the chain rule of differentiation: 
%Note: the first factor in the chain can be obtained either by differentiating ds wrt Ds and then Ds wrt D, or directly by noting that ds is the same as d at the points where it is defined and differentiating ds wrt D. These two porcedures give the same results (in the first case, there is a factor \Delta_x \Delta_y from the inverse DTFT whih is cancelled by the inverse of the same factor from the relation between Ds and D.)
%discarded $\Delta q(\ve{r})=\sum_{l=1}^{N_m}\frac{\partial \overline{q}(E)}{\partial A_l}\Delta A_l(\ve{r})$ 
\begin{align}
  \frac{\partial \overline{d}^s_{\ve{n},k}}{\partial \tilde{A}_l(\ve{u})}= \ch{\int_{\mathbb{R}^2}} \int_0^{E_{\mathrm{max}}} \frac{\partial \overline{d}^s_{\ve{n},k}}{\partial \overline{D}_k(\ve{u'})} \frac{\partial \overline{D}_k(\ve{u'})}{\partial \overline{Q}(\ve{u'},E)} \frac{\partial \overline{Q}(\ve{u'},E)}{\partial \tilde{A}_l(\ve{u})}\mathrm{d}E \ch{\mathrm{d}\mathrm{\ve{u'}}}\\ \nonumber
  =\Delta_x\Delta_y \int_0^{E_{\mathrm{max}}} e^{2\pi i\ve{u}\cdot \ve{r_n}} H_k(\ve{u},E) \frac{\partial \overline{q}(E)}{\partial A_l}\mathrm{d}E
  = \Delta_x\Delta_y\frac{1}{\overline{q}^{\mathrm{tot}}} \frac{\partial\overline{q}^{\mathrm{tot}}}{\partial A_l} H^{\mathcal{B}}_{k,l}(\ve{u}) e^{2\pi i\ve{u}\cdot \ve{r_n}}\nonumber.
\end{align}

\ch{where the delta function has allowed us to eliminate the integral over $\ve{u'}$.} Note that the conjugate symmetry of $H^{\mathcal{B}}_{k,l}(\ve{u})$ implies that $\frac{\partial \overline{d}^s_{\ve{n},k}}{\partial \tilde{A}_l(\ve{u})^*}=\left(\frac{\partial \overline{d}^s_{\ve{n},k}}{\partial \tilde{A}_l(\ve{u})}\right)^*=\Delta_x\Delta_y\frac{1}{\overline{q}^{\mathrm{tot}}} \frac{\partial\overline{q}^{\mathrm{tot}}}{\partial A_l} H^{\mathcal{B}}_{k,l}(\ve{-u}) e^{-2\pi i\ve{u}\cdot \ve{r_n}}$.

%This derivation is more extensive and has been replaced by reasoning about a linearized forward model
% \begin{align}
% \frac{\partial \overline{d}^s_{\ve{n},k}}{\partial \tilde{A}_l(\ve{u})}= \Delta_x\Delta_y\int_0^{E_{\mathrm{max}}} \int_\Rtwo\int_\Rtwo \frac{\partial \overline{d}^s_{\ve{n},k}}{\partial \overline{D}_k(\ve{u}')} \frac{\partial \overline{D}_k(\ve{u}')}{\partial \overline{Q}(\ve{u}',E)} \frac{\partial \overline{Q}(\ve{u}',E)}{\partial \overline{q}(\ve{r},E)}\frac{\partial \overline{q}(\ve{r},E)}{\partial A_l(\ve{r})}\\ \nonumber
% \cdot\frac{\partial A_l(\ve{r})}{\partial \tilde{A}_l(\ve{u})}\mathrm{d}E\mathrm{d}\ve{r}\mathrm{d}\ve{u}'\\ \nonumber
% =\Delta_x\Delta_y \int_0^{E_{\mathrm{max}}} \int_\Rtwo\int_\Rtwo e^{2\pi i\ve{u}'\cdot \ve{r_n}} H_k(\ve{u}',E) e^{-2\pi i\ve{u}'\cdot\ve{r}}\frac{\partial \overline{q}(\ve{r},E)}{\partial A_l(\ve{r})}e^{2\pi i\ve{u}\cdot\ve{r}}\mathrm{d}E\mathrm{d}\ve{r}\mathrm{d}\ve{u}'.\\ \nonumber
% \end{align}
% 
% %Superfluous text: where we have used the fact that the matrix element for the inverse fourier transform is the complex conjugate of the matrix element of the forward transform. 
% 
% By once again making the small-signal approximation, we can assume that $\frac{\partial \overline{q}(\ve{r},E)}{\partial A_l(\ve{r})}=\frac{\partial \overline{q}(E)}{\partial A_l}$ independent of position, where the derivative is evaluated at $A=A_l^0$. This causes the integrals over $x$ and $y$ to collapse:

Inserted into (\ref{eq:fisher_first_term}) this gives the elements of $\ma{F}_{\tilde{\ve{A}}}$ as
\begin{align}
  \left(\ma{F}_{\tilde{\ve{A}}}\right)_{l,l'}(\ve{u},\ve{u}')=\sum_{\ve{n}=-\infty}^{\infty} \sum_{\ve{n}'=-\infty}^{\infty} \sum_k \sum_{k'} \left(\frac{\partial \overline{d}^s_{\ve{n},k}}{\partial \tilde{A}_l(\ve{u})}\right)^*
  \left[\mathrm{Cov}(\ve{d}^s)^{-1}\right]_{\ve{n},\ve{n}',k,k'} \frac{\partial \overline{d}^s_{\ve{n}',k'}}{\partial \tilde{A}_{l'}(\ve{u}')} \\ \nonumber
  %=\sum_{\ve{n}=-\infty}^{\infty} \sum_{\ve{n}'=-\infty}^{\infty} \sum_k \sum_{k'} H^*_{k,l}(\ve{u}) e^{-2\pi i\ve{u}\cdot \ve{r_n}}\\ \nonmuber  %discarded intermediate step
  %\cdot \left[\mathrm{Cov}(\ve{d}^s)^{-1}\right]_{\ve{n},\ve{n}',k,k'} e^{2\pi i\ve{u}'\cdot \ve{n}'} H_{k',l'}(\ve{u}')
  =\frac{\Delta_x^2\Delta_y^2}{{\overline{q}^\mathrm{tot}}^2} \frac{\partial\overline{q}^{\mathrm{tot}}}{\partial A_l} \frac{\partial\overline{q}^{\mathrm{tot}}}{\partial A_{l'}} \sum_k \sum_{k'} H^{\mathcal{B}}_{k,l}(\ve{u})^* \sum_{\ve{n}=-\infty}^{\infty} \sum_{\ve{n}'=-\infty}^{\infty} e^{-2\pi i\ve{u}\cdot \ve{r_n}} 
  \left[\mathrm{Cov}(\ve{d}^s)^{-1}\right]_{\ve{n},\ve{n}',k,k'} e^{2\pi i\ve{u}'\cdot \ve{r}_{\ve{n}'}} H^{\mathcal{B}}_{k',l'}(\ve{u}')\\ \nonumber
  =\frac{\Delta_x^2\Delta_y^2}{{\overline{q}^\mathrm{tot}}^2} \frac{\partial\overline{q}^{\mathrm{tot}}}{\partial A_l}\frac{\partial\overline{q}^{\mathrm{tot}}}{\partial A_{l'}}\left[\ma{H}^{\mathcal{B}\dagger}\fourop_{\mathrm{DS}}\mathrm{Cov}(\ve{d}^s)^{-1}\fourop_{\mathrm{DS}}^\dagger\ma{H}^{\mathcal{B}}\right]_{l,l'}(\ve{u},\ve{u}') \\ \nonumber
  =\frac{1}{{\overline{q}^\mathrm{tot}}^2} \frac{\partial\overline{q}^{\mathrm{tot}}}{\partial A_l}\frac{\partial\overline{q}^{\mathrm{tot}}}{\partial A_{l'}}\left[\ma{H}^{\mathcal{B}\dagger}\left(\fourop_{\mathrm{DS}}\mathrm{Cov}(\ve{d}^s)\fourop_{\mathrm{DS}}^\dagger\right)^{-1}\ma{H^{\mathcal{B}}}\right]_{l,l'}(\ve{u},\ve{u}'),
\end{align}
where $\ma{H}^\mathcal{B}$ is the transfer matrix matrix with elements $H^{\mathcal{B}}_{k,l}(\ve{u})$. Similarly, the values of $\tilde{\ma{F}}_{\tilde{\ve{A}}}$ are given by
\begin{equation}
  \left(\tilde{\ma{F}}_{\tilde{\ve{A}}}\right)_{l,l'}(\ve{u},\ve{u}')=\frac{1}{{\overline{q}^\mathrm{tot}}^2} \frac{\partial\overline{q}^{\mathrm{tot}}}{\partial A_l}\frac{\partial\overline{q}^{\mathrm{tot}}}{\partial A_{l'}}\left[\ma{H}^{\mathcal{B}\dagger}\left(\fourop_{\mathrm{DS}}\mathrm{Cov}(\ve{d}^s)\fourop_{\mathrm{DS}}^\dagger\right)^{-1}\ma{H}^{\mathcal{B}}\right]_{l,l'}(\ve{u},-\ve{u}').
\end{equation}

Using (\ref{eq:appendix_fourier_cov_components}), we obtain the Fisher matrix elements as
\begin{align}
  \label{eq:appendix_fisher_matrix_components}
  \left(\ma{F}_{\tilde{\ve{A}}}\right)_{l,l'}(\ve{u},\ve{u}')= \frac{1}{{\overline{q}^\mathrm{tot}}^2} \frac{\partial\overline{q}^{\mathrm{tot}}}{\partial A_l}\frac{\partial\overline{q}^{\mathrm{tot}}}{\partial A_{l'}}\sum_k \sum_{k'} H^{\mathcal{B}}_{k,l}(\ve{u})^* \delta(\ve{u}-\ve{u}') \left[W_{d^+}^{-1}\right]_{k,k'}(-\ve{u}) H^{\mathcal{B}}_{k',l'}(\ve{u})\\ \nonumber
  = \frac{1}{{\overline{q}^\mathrm{tot}}^2} \frac{\partial\overline{q}^{\mathrm{tot}}}{\partial A_l}\frac{\partial\overline{q}^{\mathrm{tot}}}{\partial A_{l'}}\delta(\ve{u}-\ve{u}') \mathrm{NEQ}_{l,l'}^{\mathcal{B}}(\ve{u}).
\end{align}

%WRONG, discarded
% \begin{equation}
%   \label{eq:appendix_compl_fisher_matrix_components}
%   \left(\tilde{\ma{F}}_{\tilde{\ve{A}}}\right)_{l,l'}(\ve{u},\ve{u}')= \frac{1}{{\overline{q}^\mathrm{tot}}^2} \frac{\partial\overline{q}^{\mathrm{tot}}}{\partial A_l}\frac{\partial\overline{q}^{\mathrm{tot}}}{\partial A_{l'}}\delta(\ve{u}+\ve{u}') \mathrm{NEQ}_{l,l'}^{\mathcal{B}}(-\ve{u}).
% \end{equation}
A similar calculation shows that $\left(\ch{\tilde{\ma{F}}}_{\tilde{\ve{A}}}\right)_{l,l'}(\ve{u},\ve{u}')$ is proportional to $\delta(\ve{u}+\ve{u'})$. Since we restrict ourselves to estimating the Fourier coefficient in the right half-plane $\left\{\ve{u}: u > 0 \;\mathrm{or}\; u=0,v>0\right\}$, $\left(\tilde{\ma{F}}_{\tilde{\ve{A}}}\right)_{l,l'}(\ve{u},\ve{u}')=\ma{0}$, which gives

\begin{equation}
  \label{eq:CRLB_for_underline_A}
  \ch{\mathrm{Cov}\left(\underline{\hat{\tilde{\ve{A}}}}\right)^{-1}\le\underline{\ma{F}}_{\tilde{\ve{A}}}=
  \begin{pmatrix}
    \ma{F}_{\tilde{\ve{A}}} & \ma{0}\\
    \ma{0} & \ma{F}_{\tilde{\ve{A}}}^*
  \end{pmatrix},}
\end{equation}
\ch{with components of $\ma{F}_{\tilde{\ve{A}}}$ given by (\ref{eq:appendix_fisher_matrix_components}).}

The case $\ve{u}=\ve{0}$ requires special treatment since the fact that $\mathrm{Im}\;\tilde{\ve{A}}(\ve{0})=\ve{0}$ makes $\ch{\ma{F}}_{\tilde{\ve{A}}}$ singular if $\ve{u}=\ve{0}$ is included. However, (\ref{eq:appendix_fisher_matrix_components}) is sufficient for our purposes since $\mathrm{Cov}\left(\hat{\tilde{A}}_l(\ve{u})\right)$ can be extrapolated to $\ve{u}=\ve{0}$ as long as the NEQ \ch{is a smooth function of $\ve{u}$ near the origin.}

%does not have a squared delta function at the origin.

%Obsolete
% \begin{align}
%   \ch{\left(\ma{F}_{\tilde{\ve{A}}}\right)_{l,l'}(\ve{u},\ve{u}') = \frac{1}{{\overline{q}^\mathrm{tot}}^2} \frac{\partial\overline{q}^{\mathrm{tot}}}{\partial A_l}\frac{\partial\overline{q}^{\mathrm{tot}}}{\partial A_{l'}}\delta(\ve{u}-\ve{u}') \mathrm{NEQ}_{l,l'}^{\mathcal{B}}(\ve{u}).}
% \end{align}

%NOTE: about dimensions. Cov(\tilde{A}) has dimension length^6. Fisher(A) has dimenstion length^-2, NOT length^-6. Reason: only the diagonal elements, and not the delta function, are inverted when inverting a matrix. 

\ch{Since different spatial frequencies are independent, the Fisher information matrix $\ma{F}_{\tilde{\ve{A}}}$ is block diagonal and the block corresponding to spatial frequency $\ve{u}$ is given by a rescaled version of $\ma{NEQ}^{\mathcal{B}}(\ve{u})$, where $\mathcal{B}$ is the set of basis functions in which $\tilde{\ve{A}}$ is expressed. $\ma{F}_{\tilde{\ve{A}}}$ can thus be inverted separately for each $\ve{u}$.  This allows the CRLB (\ref{eq:CRLB_for_underline_A}) to be expressed as }

\begin{equation}
  \label{eq:CRLB_and_NEQ_appendix}
  \ch{\mathrm{Cov}\left(\hat{\tilde{A}}_l(\ve{u}),\hat{\tilde{A}}_{l'}(\ve{u'})\right) \ge {{\overline{q}^\mathrm{tot}}^2}\frac{\left[\ma{NEQ}^{\mathcal{B}}(\ve{u})^{-1}\right]_{l,l'}}{ \frac{\partial\overline{q}^{\mathrm{tot}}}{\partial A_l}\frac{\partial\overline{q}^{\mathrm{tot}}}{\partial A_{l'}}}\delta(\ve{u}-\ve{u'}).}
  %Obsolete: \mathrm{Cov}\left(\hat{\tilde{\ve{A}}}^{\mathrm{norm}}(\ve{u})\right) \ge \left[\ma{NEQ}^{\mathcal{B}}(\ve{u})\right]^{-1}
\end{equation}
\ch{which is (\ref{eq:CRLB_and_NEQ}).}
%New citations rajbhandary_frequency_dependent_dqe, 2 on image recon

\section{\ch{Detectability in a combination of basis images}}
\label{sec:appendix_CRLB_material_images}
\ch{In this section we derive the upper limit of detectability in a frequency-dependent weighted linear combination of basis images. Combining the CRLB for $\bm{\uptheta}$, $\mathrm{Cov}(\bm{\hat{\uptheta}}) \ge \ma{F_{\bm{\uptheta}}}^{-1}$, with the formula for the ideal linear observer (Sec. 13.2.12 of Ref. \citenum{barrett_myers_foundations}),}
\begin{align}
  \label{eq:d2_combination_of_basis_images}
  \ch{d'^2 \le \ve{\Delta} \bm{\uptheta} \trsp \ma{F}_{\bm{\uptheta}} \ve{\Delta} \bm{\uptheta}=
  \frac{1}{4}\ve{\Delta} \bm{\uptheta} \trsp \ma{T^\dagger}\ma{T} \ma{F}_{\bm{\uptheta}} \ma{T^\dagger}\ma{T} \ve{\Delta} \bm{\uptheta}=}
  \ch{\ve{\Delta} \underline{\tilde{\ve{A}}}^\dagger \ma{\underline{F}}_{\tilde{\ve{A}}} \ve{\Delta} \underline{\tilde{\ve{A}}}=}\\ \nonumber
  \ch{\ve{\Delta}\tilde{\ve{A}}^\dagger \ma{F}_{{\tilde{\ve{A}}}} \ve{\Delta} \tilde{\ve{A}} +\ve{\Delta}\tilde{\ve{A}}\trsp \ma{F}_{{\tilde{\ve{A}}}}^* \ve{\Delta} \tilde{\ve{A}}^*.}
  %= 2 \ve{\Delta}\tilde{\ve{A}}^\dagger \ma{F}_{{\tilde{\ve{A}}}} \ve{\Delta} \tilde{\ve{A}}\nonumber
\end{align}

\ch{The block-diagonality of the Fisher information matrix allows us to treat different spatial frequencies independently, and therefore the first term in this sum is an integral over the half-plane $u>0$. Since $\Delta \tilde{A}_{l}(\ve{u})$ and $\mathrm{NEQ}^{\mathcal{B}}(\ve{u})$ are both conjugate-symmetric, including the second term is equivalent to extending the integral to the entire $\ve{u}$ plane:}
%NOTE: the two terms are actually equal since the Fisher matrix is Hermitian
\begin{align}
  \ch{d'^2 \le\frac{1}{{\overline{q}^\mathrm{tot}}^2} \int_\Rtwo \sum_{l}\sum_{l'}\frac{\partial\overline{q}^{\mathrm{tot}}}{\partial A_l}\Delta \tilde{A}_{l}(\ve{u})^\dagger \left[\mathrm{NEQ}^{\mathcal{B}}(\ve{u})\right]_{l,l'} \Delta \tilde{A}_{l'}(\ve{u}) \frac{\partial\overline{q}^{\mathrm{tot}}}{\partial A_{l'}} \mathrm{d}\ve{u}} \\ \ch{\nonumber = \int_\Rtwo \ve{\Delta S}^{\mathcal{B}}(\ve{u})^\dagger \ma{NEQ}^{\mathcal{B}}(\ve{u}) \ve{\Delta S}^{\mathcal{B}}(\ve{u}) \mathrm{d}\ve{u}.}
\end{align}
\ch{which is (\ref{eq:d2_upper_limit_weighted_basis_image}).}

% \bibliographystyle{aipnum4-1_mph}
% \bibliography{dqe_paper}

%merlin.mbs aipnum4-1.bst 2010-07-25 4.21a (PWD, AO, DPC) hacked
%Control: key (0)
%Control: author (8) initials jnrlst
%Control: editor formatted (1) identically to author
%Control: production of article title (1) required
%Control: page (1) range
%Control: year (1) truncated
%Control: production of eprint (0) enabled
%
\clearpage
\begin{longtable}{lll}
  %\begin{tabular}{lll}
    \caption{List of symbols and their dimensions. $\mathrm{L}$: length; $\mathrm{E}$: energy.}\\
    \hline
    \hline
    Symbol & Dimension & Description\\
    \hline
    ${}^*$				&					& Complex conjugate\\
    ${}^\dagger$			&					& Conjugate-transpose\\
    $A$ 				& $\mathrm{L}$ 				& Basis coefficient line integral\\
    $\tilde{A}$ 			& $\mathrm{L}^3$ 			& Fourier transform of $A$\\
    $d$ 				& $1$ 					& Presampling detected signal\\
    $D$	 				& $\mathrm{L}^2$ 			& Fourier transform of $d$\\
    $d^s$ 				& $1$ 					& Sampled signal as discrete sequence\\
    $D^s$	 			& $1$ 					& Discrete-space Fourier transform of $d^s$\\
    $d^+$	 			& $\mathrm{L}^{-2}$ 			& Sampled signal as delta-pulse train\\
    $d'$ 				& $1$ 					& Optimal-linear-observer detectability\\
    $\ma{DQE}$				& $\mathrm{E}^{-1}$			& Detective quantum efficiency matrix in energy basis \\
    $\ma{DQE}^{\mathcal{B}}$		& $1$ 					& Detective quantum efficiency matrix in material basis\\
    $\Delta_x,\Delta_y$ 		& $\mathrm{L}$ 				& Detector pixel width and height \\
    $E$ 				& $\mathrm{E}$ 				& Incident energy of photon \\
    $\varepsilon$ 			& $\mathrm{E}$ 				& Registered energy in event \\
    $\ma{F}_{\tilde{\ve{A}}}$		& $\mathrm{L}^6$			& Fisher information matrix for $\tilde{A}$\\
    $\mathcal{F}$ 			& 					& Continuous Fourier transformation \\
    $\mathcal{F}_{DS}$ 			& 					& Discrete-space Fourier transformation \\
    $f$					& $\mathrm{L}^{-1}$			& Material basis function \\
    $\overline{G}$ 			& $1$					& Large-area gain \\
    $h$ 				& $\mathrm{L}^{-2}$ 			& Point-spread function \\
    $H$ 				& $1$ 					& Transfer function \\
    $H^\mathcal{B}$ 			& \chtwo{$\mathrm{L}^{-2}$} 		& Rescaled transfer matrix in material basis\\
    $K_d$ 				& $1$ 					& Cross-covariance matrix of $d$\\
    $K_{d^+}$ 				& $\mathrm{L}^{-4}$ 			& Cross-covariance matrix of $d^+$\\    
    $K^s$ 				& $1$ 					& Cross-covariance matrix of $d^s$\\ 
    $L$					& $1$					& Basis transformation matrix element \\
    $\mathrm{MTF_{pre}}$		& $1$					& Presampling modulation transfer function \\
    $\ve{n}=(n_x,n_y)$ 			& 1		 			& Discrete pixel coordinate \\
    $\ma{NEQ}$				& $\mathrm{L}^{-2} \mathrm{E}^{-2}$	& Noise-equivalent quanta matrix in energy basis \\
    $\ma{NEQ}^{\mathrm{B}}$		& $\mathrm{L}^{-2}$ 			& Noise-equivalent quanta matrix in material basis \\
    $\mathrm{Nyq}$ 			&  		 			& Nyquist region of the (u,v) plane \\
    $\Phi$				& $\mathrm{L}^{-2} \mathrm{E}^{-1}$	& Prepatient photons per area and energy\\
    $q$ 				& $\mathrm{L}^{-2} \mathrm{E}^{-1}$ 	& Incident photons per area and energy\\
    $\overline{q}^{\mathrm{tot}}$ 	& $\mathrm{L}^{-2}$ 			& Total incident photons per area\\    
    $Q$ 				& $\mathrm{E}^{-1}$ 			& Fourier transform of $q$\\    
    $\ve{r}=(x,y)$ 			& $\mathrm{L}$ 				& Position on detector\\
    $\ve{r_n}$ 				& $\mathrm{L}$ 				& Pixel center position\\
    $\ve{\Delta S}$			& $\mathrm{L}^2$ 			& Task function in energy basis\\
    $\ve{\Delta S}^\mathcal{B}$		& $\mathrm{L}^2$			& Task function in material basis\\
    $\ma{T}$				& $1$					& Matrix mapping $\ve{X}$ to $\left(\ve{X},\ve{X}^*\right)\trsp$\\
    $\ve{u}=(u,v)$ 			& $\mathrm{L}^{-1}$ 			& Spatial frequency \\
    $W^s$				& $1$					& Cross-spectral density of $d^s$\\
    $W_{d^+}$				& $\mathrm{L}^{-2}$			& Cross-spectral density of $d^+$\\
    $\overline{X}$			& 					& Expected value of $X$\\
    \hline
    \hline
  %\end{tabular}
  \label{table:symbols}
\end{longtable}
%Omitted: M, A tilde hat, T, I, underline, mu mono, transpose, q0. Have not distinguished the two things named H. Double-check dimension of F, L and Delta S
\clearpage

%\clearpage

%NOTE CRLB and CRLBrel have different signs for the off-diagonal elements since the rescaling factor is negative. I don't want to change the definitions as Delta A and Delta S should have opposite signs.

\end{document}